\documentclass[12pt]{article}
\usepackage{geometry}
\usepackage{graphicx} 
\usepackage{float}
\usepackage{nameref}  
\usepackage{amsmath}
\usepackage{authblk}
\usepackage{comment}
\usepackage{xcolor}
\usepackage{url}
\usepackage[T1]{fontenc}

\usepackage{rotating}  
\usepackage{booktabs}  
\usepackage{multirow}  

\title{Spectral Sampling of Boron Diffusion in Ni Alloys: Cr and Mo Effects on Bulk and Grain Boundary Transport}

\author[1,2]{Tyler D. Dole\v{z}al\footnote{corresponding authors: tyler.dolezal.1@us.af.mil; rodrigof@mit.edu; liju@mit.edu}}
\author[1]{Rodrigo Freitas$^*$}
\author[1,3]{Ju Li$^*$}
\affil[1]{Department of Materials Science and Engineering, Massachusetts Institute of Technology, Cambridge, MA, USA}
\affil[2]{Department of Engineering Physics, Air Force Institute of Technology, Wright-Patterson Air Force Base, OH, USA}
\affil[3]{Department of Nuclear Science and Engineering, Massachusetts Institute of Technology, Cambridge, MA, USA}

\begin{document}

\maketitle

\begin{abstract}
\noindent
Understanding how light interstitials migrate in chemically complex alloys is essential for predicting defect dynamics and long-term stability. Here, we introduce a spectral sampling framework to quantify boron diffusion activation energies in Ni and demonstrate how substitutional solutes (Cr, Mo) reshape interstitial point defect transport in both the bulk and along crystallographic defects. In the bulk, boron migration energy distributions exhibit distinct modality tied to solute identity and spatial arrangement: both Cr and Mo raise barriers in symmetric cages but induce directional asymmetry in partially decorated environments. Extending this framework to a $\Sigma5\langle100\rangle{210}$ symmetric tilt grain boundary reveals solute-specific confinement effects. Cr preserves low-barrier in-plane mobility while suppressing out-of-plane transport, guiding boron into favorable midplane voids. Mo, by contrast, imposes an across-the-board reduction in boron mobility, suppressing average diffusivity by two additional orders of magnitude at 800 $^\circ$C and reducing out-of-plane transport by five orders of magnitude relative to Cr. Both elements promote segregation by producing negative segregation energies, but their roles diverge: Cr facilitates rapid redistribution and stabilization at interfacial sites, consistent with Cr-rich boride formation, while Mo creates deeper and more uniform segregation wells that strongly anchor boron. Together, these complementary behaviors explain the experimental prevalence of Cr- and Mo-rich borides at grain boundaries and carbide interfaces in Ni-based superalloys. More broadly, we establish spectral sampling as a transferable framework for interpreting diffusion in disordered alloys and for designing dopant strategies that control transport across complex interfaces.
\end{abstract}

\section*{Introduction}

The migration of light interstitials through metal lattices plays a central role in microstructural evolution, defect chemistry, and performance degradation in structural alloys \cite{mehrerDiffusionSolids2007, simsSuperalloysIIHighTemperature1987}. In Ni-based systems, interstitial species such as boron and carbon are known to influence a wide range of properties, including grain boundary cohesion, creep resistance, oxidation tolerance, and susceptibility to embrittlement \cite{simsSuperalloysIIHighTemperature1987}. While their macroscopic effects have been widely documented \cite{bashirEffectInterstitialContent1993a, tianSynergisticEffectsBoron2024b, gongMicrostructuralEvolutionMechanical2023a, garosshenEffectsZrStructure1987, zhangSynergyPhaseMC2024, liInfluenceCarbidesPores2024a}, the mechanisms by which interstitials navigate the chemically and structurally heterogeneous landscapes of real alloys remains an active area of research.

Experimental and computational studies have provided valuable insights into both bulk and interfacial diffusion. Tracer-based measurements have revealed alloy-specific trends and the impact of phase composition on boron mobility~\cite{wangDiffusionBoronAlloys1995, keddamAssessmentBoronDiffusivities2022}. First-principles calculations have clarified site preferences and migration pathways for various interstitial species~\cite{davidFirstprinciplesStudyInsertion2020a, epifanoFirstprincipleStudySolubility2020, zhangFirstPrinciplesCalculation2018}, often treating the host lattice as chemically and structurally uniform. Other studies have introduced substitutional dopants to explore how solute identity affects interstitial migration through lattice distortion or orbital interactions~\cite{jiEffectRefractoryElements2023a}. In parallel, a growing body of work has examined how extended defects such as grain boundaries alter interstitial behavior. Rajeshwari et al.~\cite{rajeshwarik.GrainBoundaryDiffusion2020} linked GB structure and complexion transitions to changes in diffusivity. Yang et al.~\cite{yangFirstprinciplesInvestigationInteraction2017a} and Di Stefano et al.~\cite{distefanoFirstprinciplesInvestigationHydrogen2015} found that the local atomic packing of symmetric and asymmetric $\Sigma$ boundaries governs their ability to trap or transmit interstitials. He et al.~\cite{heFirstprinciplesStudyHydrogen2021}, Huang et al.~\cite{huangHydrogenAtomSolution2023}, and Dong et al.~\cite{dongFastHydrogenDiffusion2017} showed that certain GBs can serve as either barriers or high-diffusivity channels depending on crystallographic character and impurity decoration. These results point to a dual dependence of interstitial migration on both chemical coordination and defect topology. Previous studies isolate either a fixed geometry with varied chemistry~\cite{jiEffectRefractoryElements2023a}, or a fixed chemistry across different GBs~\cite{yangFirstprinciplesInvestigationInteraction2017a, heFirstprinciplesStudyHydrogen2021}, leaving open the question of how chemical heterogeneity modulates diffusion within specific interfacial structures. To address this, recent work by Xing et al.~\cite{xingNeuralNetworkKinetics2024} introduced a neural network kinetics model capable of quantifying the full spectrum of vacancy migration barriers in chemically complex alloys. By leveraging local atomic descriptors and stochastic sampling, their approach revealed how site-to-site variability in chemical environment governs kinetic activation energies, leading to a rugged migration landscape even within a single phase. Importantly, this framework emphasizes that transport in disordered alloys is governed not by a single representative barrier, but by a distribution of pathways whose accessibility depends on local atomic identity and arrangement.

Building on the spectral migration barrier analysis of Xing et al.~\cite{xingNeuralNetworkKinetics2024, xingShortrangeOrderLocalizing2022}, we extend this approach to interstitial diffusion and show how substitutional chemistry shapes point defect transport in both the bulk and along crystallographic defects. In Ni-based superalloys, boron is a crucial micro-alloying element that promotes high-temperature mechanical performance, exhibiting a strong and recurring association with Cr- and Mo-rich environments. Atom probe and electron microscopy studies consistently show boron preferentially localized within Cr--Mo-based grain boundary borides, co-segregated with Cr and Mo at random high-angle grain boundaries, and enriched at Cr--Mo carbide interfaces, while remaining largely absent from chemically incompatible interfaces such as coherent twin boundaries or $\gamma/\gamma'$ interfaces~\cite{cadelAtomProbeTomography2002, tytkoMicrostructuralEvolutionNibased2012, kontisEffectBoronGrain2016a, duPrecipitationEvolutionGrain2017a, kontisRoleBoronImproving2017a}. More recent work under additive manufacturing and service-relevant creep conditions further demonstrates that boron released from ceramic reinforcements preferentially re-partitions into (Cr,Mo)-based borides anchored at grain boundaries, where these complexes stabilize boundary structure, suppress deleterious Cr--Mo TCP phases, and correlate directly with enhanced high-temperature strength and ductility~\cite{tekogluMetalMatrixComposite2024b, tekogluSuperiorHightemperatureMechanical2024c}.

While extensive studies confirm boron's thermodynamic stability and precipitation mechanisms with Cr and Mo, the atomistic kinetic mechanisms, specifically how Cr and Mo locally perturb the potential energy landscape and alter the energy barriers and pathways for interstitial boron migration, remain poorly resolved. Importantly, while the literature on B--Cr and B--Mo ordering in Ni-based superalloys is robustly supported by experimental observation, the corresponding computational work on boron diffusion, reviewed in the previous passage, is largely limited to pure Ni simulation systems. Atomistic studies that consider boron's behavior in true alloy simulation environments are extremely limited~\cite{jiEffectRefractoryElements2023a, dolezalSegregationOrderingLight2025, dolezalAtomisticMechanismsOxidation2025, dolezalAtomisticSimulationsShortrange2025}. Additionally, standard theoretical methods often focus on an average energy barrier, masking the underlying spectrum of site-to-site energy variations essential for long-range diffusion in chemically complex alloys. 

To resolve this critical knowledge gap, we adopt a coordination-resolved spectral framework in which boron migration barriers are computed using Nudged Elastic Band (NEB) calculations across a systematically varied set of atomic neighborhoods. We consider a different perspective: interstitial diffusion as a sampling problem over a distribution of local environments. Specifically, we vary the number and arrangement of Cr or Mo atoms in the first-nearest neighbor (1NN) cage surrounding interstitial sites. This sampling approach is then extended to a grain boundary system representative of a common configuration in FCC alloys: the $\Sigma5\,\langle100\rangle\,\{210\}$ symmetric tilt boundary~\cite{schuhUniversalFeaturesGrain2005a, tschoppSymmetricAsymmetricTilt2015, zhengGrainBoundaryProperties2020}. By evaluating migration barriers in both bulk and interfacial environments, we assemble a high-resolution dataset that captures the coupled effects of local chemical coordination and grain boundary structure on interstitial transport in chemically complex Ni alloys.

\section*{Methods}

\begin{figure}[H]
    \centering
    \includegraphics[width=\linewidth]{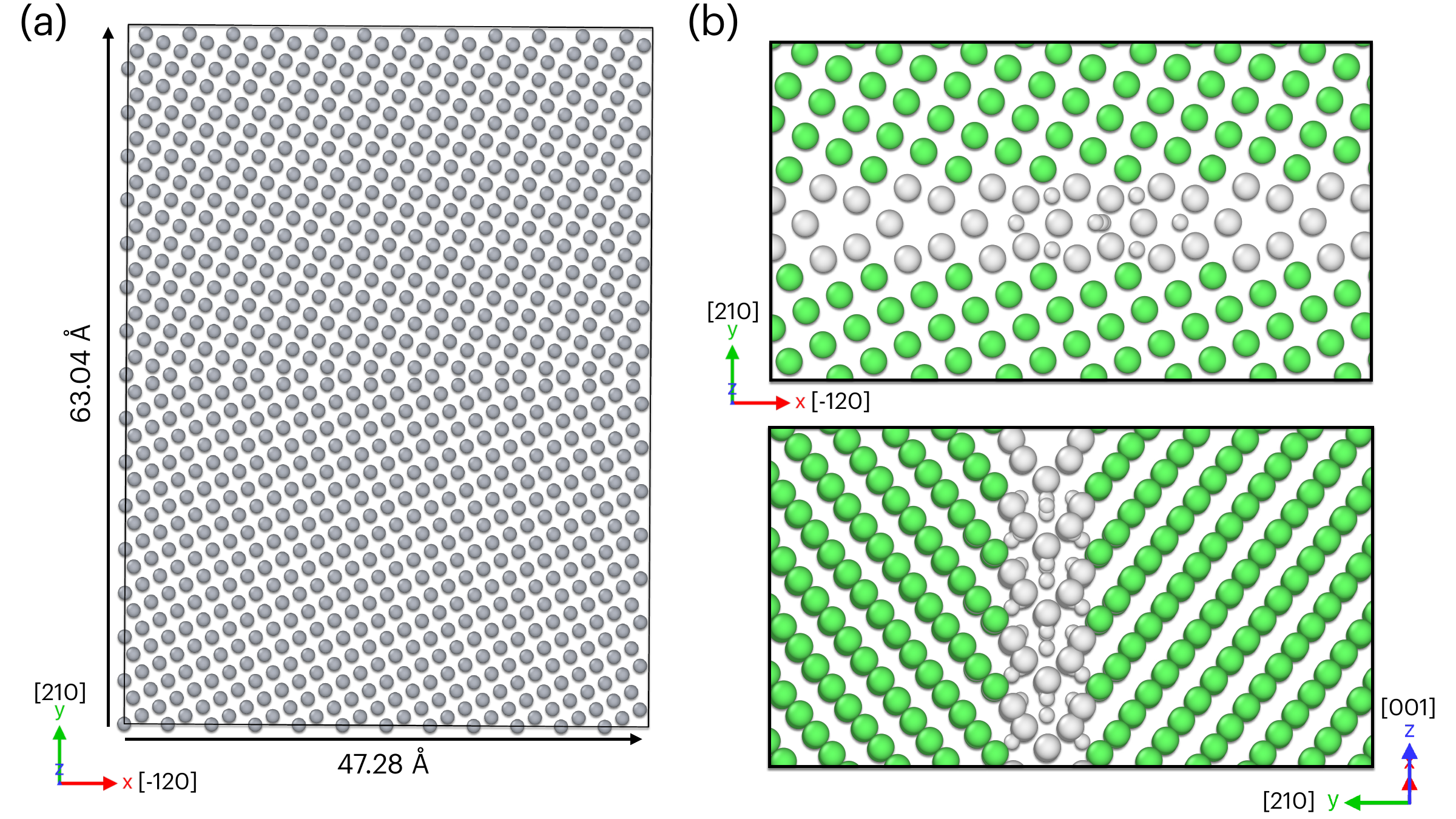}
    \caption{Visualization of the simulation cell used for grain boundary (GB) diffusion analysis, rendered in OVITO~\cite{stukowskiVisualizationAnalysisAtomistic2009a}. (a) The pure Ni $\Sigma5\,\langle100\rangle\,\{210\}$ symmetric tilt GB cell (S5) containing 4,800 atoms. The out-of-plane dimension ($\hat{z}$) measures 17.62~\AA. (b) Detailed view of the S5 interface structure with common neighbor analysis coloring. Predetermined interstitial voids are marked by the smaller atomic species. Green denotes ``fcc'' coordination, while white indicates ``other.''}
    \label{fig:simulation_cells}
\end{figure}

\subsection*{Preparing the Simulation Environment}

The bulk simulation cell was constructed as a pure Ni supercell in the face-centered cubic (FCC) structure (space group Fm$\bar{3}$m), consisting of 1,372 atoms with a lattice parameter of 24~\AA. The cell size was selected based on an analysis of the radial strain induced by boron insertion at the octahedral interstitial site, as well as the subsequent local distortion caused by Cr or Mo substitution in the first-neighbor cage (given in Fig.~S1). This ensured that the computed migration energy barriers were converged with respect to cell size (full details are provided in the Supplemental Materials). The $\Sigma5\,\langle100\rangle\,\{210\}$ grain boundary (GB) system was generated using Atomsk~\cite{hirelAtomskToolManipulating2015} and will be referred to as S5. The S5 simulation cell contained 4,800 lattice atoms. The dimensions of the cell were selected to ensure that radial strain from boron insertion and Cr or Mo substitution decayed fully within the cell. The corresponding simulation cell dimensions are provided in Fig.~\ref{fig:simulation_cells}. The calculated GB energy was $\gamma_{\mathrm{S5}} = 1.20$~J/m\textsuperscript{2}, which is consistent with literature values \cite{distefanoFirstprinciplesInvestigationHydrogen2015,zhengGrainBoundaryProperties2020, rohrerComparingCalculatedMeasured2010}.

It is important to note that while the present study focuses on a low-energy symmetric tilt boundary for its structural simplicity and suitability for systematic spectral sampling, this boundary type represents only a minority fraction of boundaries in polycrystalline Ni-based superalloys, where random high-angle boundaries dominate the network~\cite{schuhUniversalFeaturesGrain2005a,rohrerComparingCalculatedMeasured2010}. Nevertheless, the spectral sampling framework introduced here is not restricted to this GB character and can be applied to any boundary geometry, including experimentally measured networks, by repeating the same local coordination-dependent sampling protocol. Although absolute segregation energies and diffusion barriers will vary with GB structure, the qualitative mechanistic distinctions identified here are expected to persist across a broad range of boundary types.

\subsection*{Computational Details}

All simulations were performed using the Large-scale Atomic Molecular Massively Parallel Simulator (LAMMPS)~\cite{thompsonLAMMPSFlexibleSimulation2022a} in conjunction with version 5.0.0 of the universal neural network potential (Preferred Potential, PFP)~\cite{takamotoUniversalNeuralNetwork2022}, with D3 dispersion corrections applied via Matlantis~\cite{Matlantis}. Structural relaxation was carried out using conjugate gradient (CG) minimization, with an energy tolerance of 0.0~eV to ensure a force convergence of $1 \times 10^{-5}$~eV/\AA. Climbing image NEB calculations~\cite{henkelmanClimbingImageNudged2000} were performed using the Atomic Simulation Environment~\cite{hjorthlarsenAtomicSimulationEnvironment2017a}, also employing the PFP potential. Each migration pathway was discretized into six images, connected by harmonic springs with a stiffness of 0.5~eV/\AA, and relaxed to a force tolerance of 0.01~eV/\AA. The accuracy and transferability of the PFP framework have been extensively validated by its developers and in our previous works. For ease of access and transparency, we have compiled a summary of key DFT-to-PFP validation results in Tables S2 and S3 of the Supplemental Materials.

\subsection*{Spectral Sampling and Analysis}

\begin{figure}[H]
    \centering
    \includegraphics[width=\linewidth]{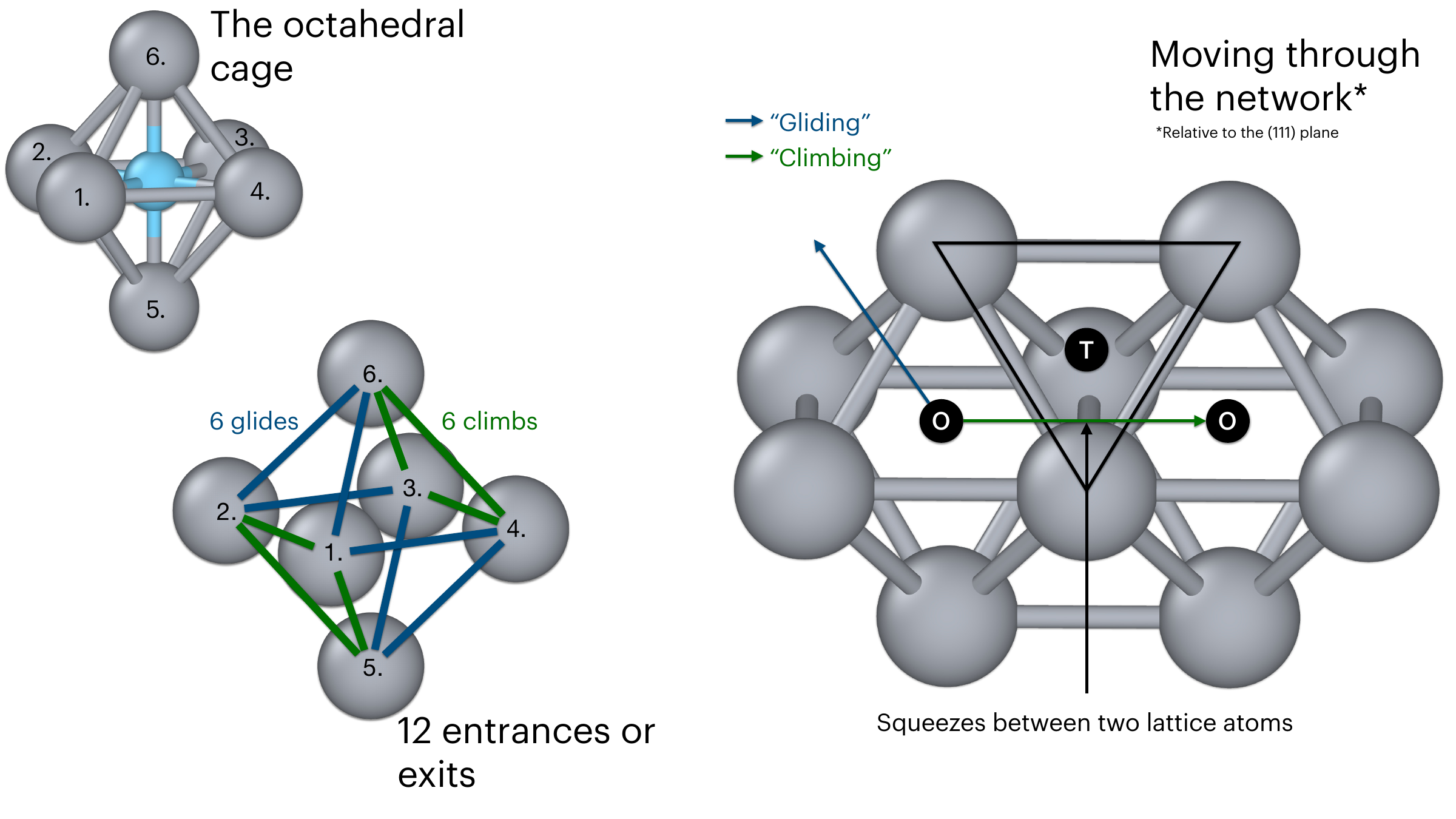}
    \caption{Schematic of the face-centered cubic octahedral cage with a boron atom positioned at the pristine octahedral interstitial site. In contrast to the conventional octahedral--tetrahedral--octahedral (O--T--O) diffusion pathway, boron migrates via one of twelve available exits leading to neighboring octahedral sites by passing near, but not through, the tetrahedral site. Instead, boron ``squeezes'' between two lattice atoms (through the edge of the tetrahedral cage) following what is commonly referred to as the \textit{direct interstitial pathway}. Depending on the direction of migration, boron can either ``glide'' (remaining within the same (111) atomic plane) or ``climb'' (moving into an adjacent (111) plane above or below the original site).}
    \label{fig:octhedral_cage}
\end{figure}

For bulk sampling, a single boron interstitial was inserted at the center of the simulation cell. Each configuration was fully relaxed, and the six nearest-neighbor Ni atoms surrounding the interstitial site were iteratively substituted with up to six Cr or Mo atoms, selected based on their proximity to the boron atom. In all cases, boron relaxed into an octahedral interstitial site. For each Cr/Mo coordination number $n$, both the initial and final octahedral sites were independently relaxed prior to performing NEB calculations. In the bulk, boron migrated via the \textit{direct interstitial pathway} \cite{mehrerDiffusionSolids2007}, as illustrated in Fig.~\ref{fig:octhedral_cage}. This pathway is in agreement with previous DFT-based studies on boron diffusion in FCC Ni \cite{jiEffectRefractoryElements2023a,yangFirstprinciplesInvestigationInteraction2017a, yangFirstprinciplesInvestigationInteraction2016a}. For comparison, the octahedral--tetrahedral--octahedral (O--T--O) diffusion pathway is described in Fig.~S3.

Interstitial migration within the GB region required a different approach due to disrupted lattice periodicity. For the S5 system, candidate interstitial sites were identified using a Voronoi-based void-finding algorithm restricted to the bounded GB zone (Fig.~\ref{fig:simulation_cells}b). Voids were filtered based on minimum radius and spatial location to ensure physically meaningful insertion sites. The resulting set of positions captured the structural and chemical heterogeneity of the GB. Migration pathways were constructed by connecting neighboring voids within a 2.5~\AA\ cutoff, producing a spatially resolved network of possible interstitial hops. The migration barrier spectrum was then computed for each unique local coordination environment, reflecting the energetic impact of both substitutional disorder and interfacial topology.

The activation energy barriers for boron diffusion in bulk Ni and the S5 GB are in good agreement with prior DFT-based studies~\cite{jiEffectRefractoryElements2023a,yangFirstprinciplesInvestigationInteraction2017a,yangFirstprinciplesInvestigationInteraction2016a}. Importantly, the values reported here were carefully converged with respect to simulation cell size. For smaller supercells (e.g., comparable in scale to those used in DFT calculations) the predicted barriers were slightly higher and aligned more closely with published DFT results. As the supercell dimensions increased, the migration barriers systematically decreased and ultimately converged to the values presented throughout this manuscript.

To quantify boron co-segregation with Cr or Mo at the S5 GB, segregation energies were computed relative to the undoped system:
\begin{equation}\label{eq:eseg}
E_\mathrm{seg}(n) = \left[ E_\mathrm{GB}(n) - E_\mathrm{GB}^{\mathrm{undoped}} \right] - \left[ \langle E_\mathrm{bulk}^{(n)} \rangle - E_\mathrm{bulk}^{\mathrm{undoped}} \right].
\end{equation}
Here, $ E_\mathrm{GB}(n) $ is the total energy of the GB system with boron coordinated by $ n $ Cr or Mo atoms, and $ E_\mathrm{GB}^{\mathrm{undoped}} $ is the total energy of the corresponding undoped GB system. $ \langle E_\mathrm{bulk}^{(n)} \rangle $ is the average energy of bulk configurations where boron is inserted with the same coordination number $ n $, and $ E_\mathrm{bulk}^{\mathrm{undoped}} $ is the energy of the undoped bulk reference. This formulation enables a consistent comparison between bulk and GB configurations. When plotted against Cr or Mo enrichment, the variability in the spectra reflects how local short-range ordering modifies the energetic landscape for light interstitial segregation. A negative $E_\mathrm{seg}(n)$ indicates that the GB configuration is more stable than its bulk counterpart at the same coordination level, signifying a favorable segregation tendency. For $n=0$, this corresponds to the segregation energy of boron in pure Ni. For $n>0$, it represents the segregation energy of the B--Cr or B--Mo motif to the GB. The reported values of boron segregation energy to the various voids within the S5 GB are in good agreement with previously published DFT results~\cite{chenEnhancingSulfurEmbrittlement2025a}.

\section*{Results and Discussion}

\subsection*{Activation Energy Spectra}

\begin{figure}[H]
    \centering
    \includegraphics[width=\linewidth]{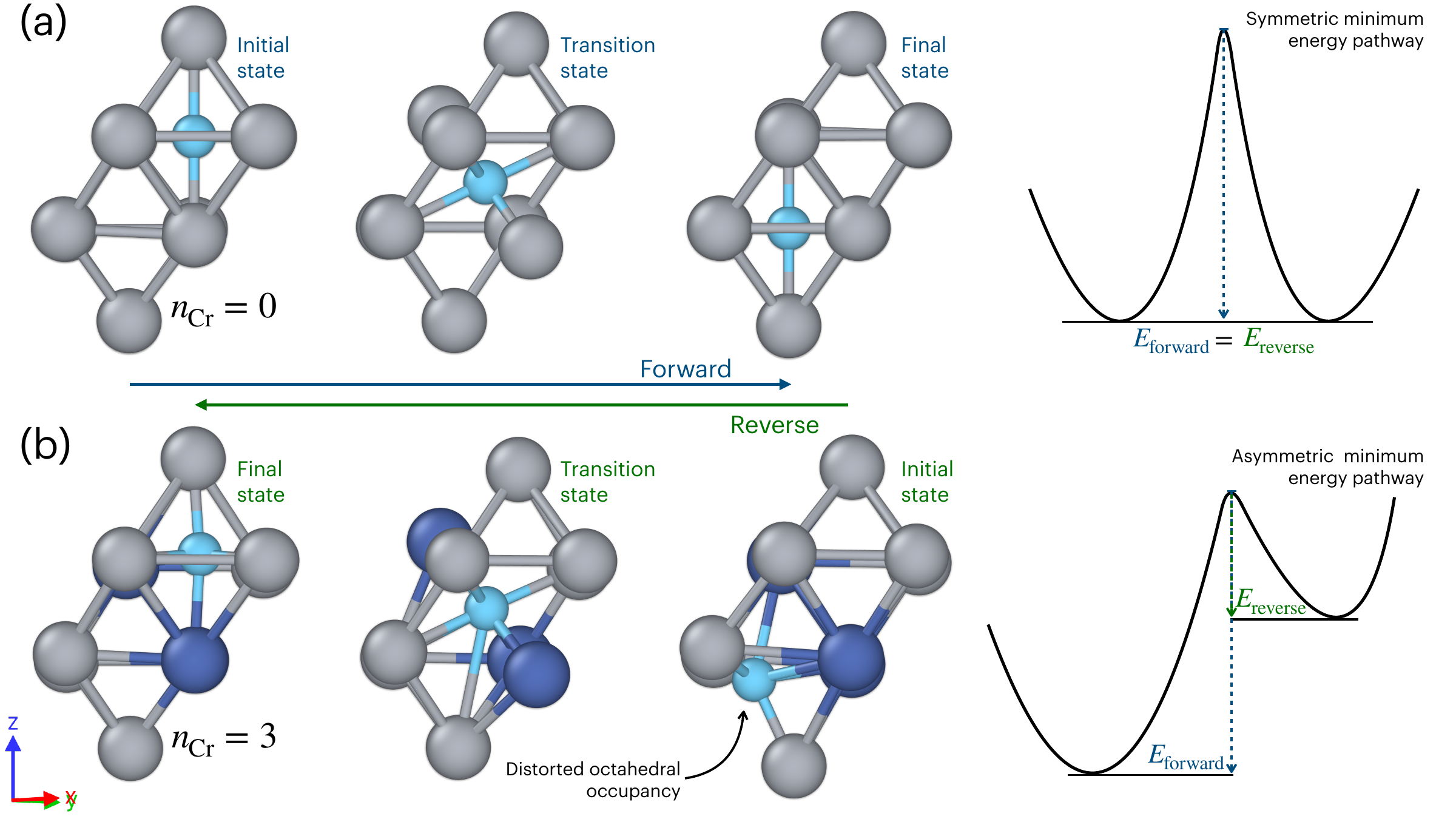}
    \caption{Schematics and corresponding minimum energy pathways for a boron bulk diffusion event. (a) illustrates the event in the pure Ni system ($n=0$ Cr), where the minimum energy pathway is symmetric ($E_{\mathrm{forward}} = E_{\mathrm{reverse}}$). (b) shows the geometric distortion and boron occupancy for the same event in the alloy at $n=3$ Cr, where the chemical decoration induces an asymmetric (or directionally biased) minimum energy pathway ($E_{\mathrm{forward}} \neq E_{\mathrm{reverse}}$).}
    \label{fig:distortion}
\end{figure}

Before conducting a detailed analysis of the spectral data, it is important to establish an intuitive understanding of the processes occurring within the simulation environments. Fig.~\ref{fig:distortion} has been provided to help anchor the discussion around the activation energy spectra. Firstly, Fig.~\ref{fig:distortion}a demonstrates the simplest boron diffusion event: migration between equivalent octahedral sites in pure Ni ($n=0$ Cr). There is no cage distortion outside of the local strain induced by boron's presence, and the result is a symmetric minimum energy pathway. Next, Fig.~\ref{fig:distortion}b demonstrates how boron diffusion is modified as a result of Cr alloying in the nearest neighbor shell ($n=3$ Cr). Due to the B--Cr interaction, the boron atom is pushed off-center, resulting in a distorted interstitial configuration. Consequently, the minimum energy pathway for this event is asymmetric, or directionally biased, as a result of the B--Cr interactions. Additionally, Fig.~\ref{fig:distortion} serves to define the directional nomenclature: walking from left-to-right represents the ``forward'' event, while walking from right-to-left yields the ``reverse'' event. As chemical coordination scales from $n=0$ to $n=6$, the forward event is effectively boron diffusing into the chemically decorated cage, and the reverse event is boron diffusing away from the chemically decorated cage. Note that adjacent octahedral cages share two metal neighbors. This geometrical constraint means that the forward initial state (and reverse final state) can be partially decorated.

\begin{figure}[H]
    \centering
    \includegraphics[width=\linewidth]{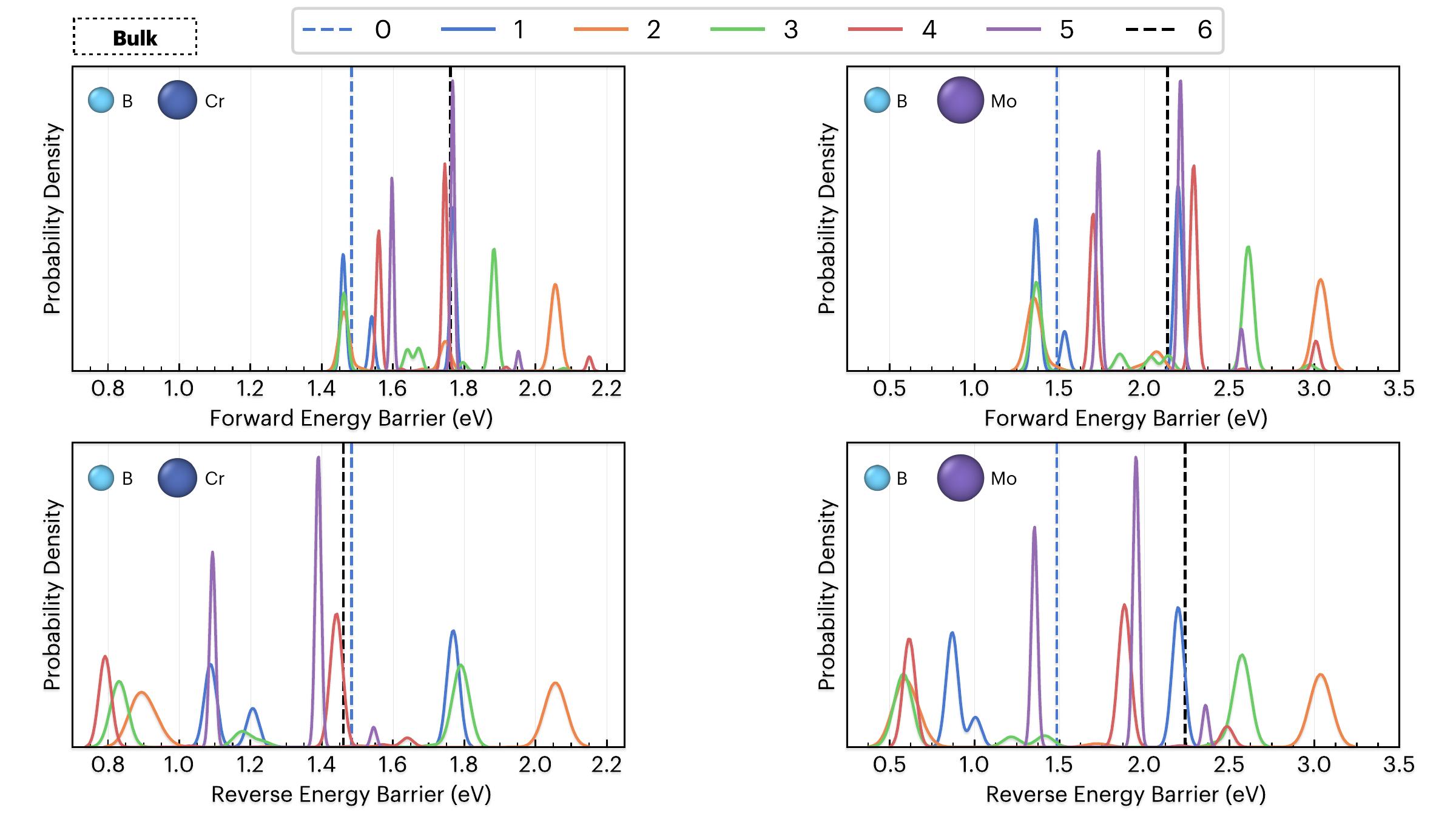}
    \caption{Activation energy spectra for boron diffusion via the \textit{direct interstitial pathway} in bulk face-centered cubic Ni. The top panel shows the forward activation energy barriers as a function of Cr and Mo coordination in the octahedral cage, while the bottom panel presents the corresponding reverse barrier distribution. The colors correspond to the number of Cr or Mo neighbors in the final state octahedral cage.}
    \label{fig:bulk_spectra}
\end{figure}

The spectra for boron diffusion in the bulk are compiled in Fig.~\ref{fig:bulk_spectra}. Peak amplitudes reflect the frequency with which distinct local motifs contribute to a given barrier height. A tall peak indicates that many sampled configurations share the same chemical and structural environment, whereas smaller peaks arise from less common motifs. Peak height should be interpreted as motif multiplicity rather than an intrinsic probability or transition rate. The observed bi- and tri-modal energy barrier distributions presented in Fig. \ref{fig:bulk_spectra} arise from distinct permutations by which Cr (or Mo) atoms decorate the octahedral cage surrounding the interstitial site. Initially, any of the six cage vertices are energetically equivalent, owing to the symmetric Ni lattice and the equidistant coordination of boron. Upon substitution, however, the boron atom tends to relax away from the nearest Cr neighbor due to the longer B--Cr bond length compared to Ni--B (Fig.~S2). When a Cr atom occupies a lattice site along the diffusion pathway, specifically one of the two edge sites through which boron enters the octahedral cage, both the forward and reverse energy barriers increase. In this configuration, Cr presents a steric and electronic obstruction that restricts boron's movement. Conversely, when Cr resides on the opposite edge of the cage, boron relaxes off-center toward the entry point, effectively reducing the migration distance and barrier height. A third scenario arises when Cr sits at a lateral or off-axis position, neither aligned with nor directly opposing the migration direction. In this case, boron's path is not fully obstructed, but the local distortion still perturbs the energy landscape. These configurations lead to intermediate barrier heights, completing the tri-modal distribution. This directional dependence applies to both the ``glide'' and ``climb'' transition routes. Similar behavior is observed with Mo substitution, although the associated energy penalties are more pronounced due to Mo's larger atomic radius, longer B--Mo bond length (Fig.~S2), and the greater stability of B--Mo interactions relative to Ni--B. Notably, the highest migration barriers for both Cr and Mo are observed at $n = 2$. This configuration corresponds to a symmetric case in which the two substitutional atoms are shared between the initial and final octahedral cages. As a result, both the starting and ending positions of the boron atom are perturbed, yielding maximal energetic resistance to migration. Asymmetric variations of this case are the higher energy $n=3$ and $n=5$ peaks. Corresponding values for all peaks have been compiled in Table \ref{tab:barrier_spectra_peaks}. The displacement of boron from the $n=0$ site as a function of Cr or Mo coordination is given in Fig.~S6 of the Supplemental Materials. These observations reveal that the heterogeneity in the migration barrier spectrum is governed by geometric positioning and the chemical identity of neighboring atoms. The resulting energy landscape is thus shaped by a complex interplay of local structure and species-specific bonding, which directly influences interstitial mobility in chemically disordered environments.

\begin{table}[H]
\centering
\begin{tabular}{c|p{0.35\linewidth}|p{0.35\linewidth}}
\toprule\toprule
$\boldsymbol{n}$ & $\boldsymbol{E_{\mathrm{forward}}}$ \textbf{ (eV)} & $\boldsymbol{E_{\mathrm{reverse}}}$ \textbf{ (eV)} \\
\midrule
\multicolumn{3}{c}{\textbf{Cr (Bulk)}} \\
\midrule
0 & 1.484 & 1.484 \\
1 & 1.460, 1.540, 1.769 & 1.086, 1.205, 1.770 \\
2 & 1.463, 1.747, 2.054 & 0.894, 2.053 \\
3 & 1.462, 1.639, 1.670, 1.884 & 0.830, 1.176, 1.790 \\
4 & 1.561, 1.746 & 0.792, 1.443, 1.640 \\
5 & 1.597, 1.767, 1.952 & 1.094, 1.391, 1.545 \\
6 & 1.761 & 1.460 \\
\midrule
\multicolumn{3}{c}{\textbf{Mo (Bulk)}} \\
\midrule
0 & 1.484 & 1.484 \\
1 & 1.364, 1.530, 2.196 & 0.871, 1.003, 2.197 \\
2 & 1.349, 2.071, 3.038 & 0.597, 3.032 \\
3 & 1.362, 1.858, 2.038, 2.137, 2.615 & 0.582, 1.219, 1.408, 2.577 \\
4 & 1.700, 2.292, 3.007 & 0.611, 1.882, 2.492 \\
5 & 1.733, 2.214, 2.569 & 1.354, 1.949, 2.359 \\
6 & 2.137 & 2.240 \\
\bottomrule\bottomrule
\end{tabular}
\caption{Peak positions (in eV) in the activation energy spectra for boron migration through Cr- and Mo-decorated octahedral cages in bulk Ni.}
\label{tab:barrier_spectra_peaks}
\end{table}

Focusing on the activation energy spectra in Fig.~\ref{fig:bulk_spectra} and Table~\ref{tab:barrier_spectra_peaks}, it is instructive to begin with the symmetric cases where $ n = 0 $ and $ n = 6 $. In the figures, these configurations are denoted by blue and black dashed lines, respectively, since they yield a single migration barrier rather than a distribution. The forward energy barrier ($ E_{\mathrm{forward}} $) in these cases reflects the energetic cost for boron to migrate from a predominantly Ni-coordinated octahedral cage into one fully decorated with Cr or Mo atoms. Owing to lattice symmetry, two atoms are always shared between adjacent cages. The introduction of Cr or Mo increases the barrier for forward diffusion, with values rising from 1.484~eV (for $ n = 0 $) to 1.761~eV (Cr, $ n = 6 $) and 2.137~eV (Mo, $ n = 6 $), respectively. However, the reverse barriers reveal a subtle but important asymmetry. For Cr, escape from the decorated cage is easier than entry, with $ E_{\mathrm{reverse}} = 1.460 $~eV for $ n = 6 $, compared to its forward barrier of 1.761~eV. In contrast, Mo effectively traps boron relative to Ni or Cr, with a reverse barrier of 2.240~eV for $ n = 6 $, exceeding even its forward value. This indicates that while both species impede entry into decorated cages, only Mo significantly inhibits escape as well.

Moving beyond the symmetric extremes, the asymmetric cases ($ 1 \leq n \leq 5 $) introduce a more complex distribution of migration barriers. These reflect the varied local geometries and chemical environments encountered during interstitial motion. Most configurations with partial decoration elevate the forward barrier, inhibiting entry into Cr- or Mo-containing cages. Yet, in several cases, the reverse barrier drops below that of the pure Ni case, suggesting that certain motifs promote boron escape. This behavior can be broadly interpreted as a leftward shift in the reverse barrier spectrum relative to the forward one; that is, many reverse events resemble their forward counterparts but with a systematically lower energy. This shift reflects the directional asymmetry imposed by the nonuniform decoration of neighboring cages. Only in cases where the chemical configuration is symmetric about the forward and reverse paths (e.g., $ n = 0 $ or $ n = 6 $) do the forward and reverse barriers coincide exactly. For example, at $ n = 3 $, Cr-decorated configurations exhibit reverse barriers as low as 0.830~eV, despite forward barriers ranging as high as 1.884~eV, indicating that certain asymmetric motifs promote easier escape than entry. A similar trend is observed for Mo, where reverse barriers reach as low as 0.582~eV at $ n = 3 $, again underscoring the energetic asymmetry. The higher reverse barrier values within the same $ n $ group, which closely match their forward counterparts, correspond to more symmetric local environments and thus serve as exceptions to the broader trend. These directional asymmetries are critical for understanding interstitial mobility in chemically complex systems where light interstitials traverse a disordered landscape of partially decorated cages. In such systems, the relative decoration of the ``gaining'' and ``losing'' cages effectively serves to either push or pull the interstitial along its trajectory.

The distinction between Cr and Mo substitutions further complicates the energetic landscape. Mo consistently induces broader and more extreme energy spreads, with forward barriers spanning 1.349~eV up to 3.038~eV and reverse barriers ranging from 0.582~eV to 3.032~eV. Notably, the upper end of this range corresponds to a nearly symmetric configuration at $ n = 2 $, where the forward and reverse pathways are chemically equivalent. Due to its larger atomic radius and longer bond length with boron (Fig.~S2), Mo also exhibits stronger directional asymmetry than Cr. In configurations that facilitate escape, Mo-decorated cages offer reverse barriers as low as 0.582~eV (at $ n = 3 $), compared to 0.830~eV for Cr. Conversely, in trapping configurations, Mo's reverse barriers far exceed those of Cr, confirming its superior ability to confine boron. Even under full decoration ($ n = 6 $), Cr slightly promotes boron escape relative to the undistorted Ni cage, while Mo more effectively suppresses it. These findings underscore how local chemical environment, coordination asymmetry, and elemental identity collectively shape the potential energy landscape for light interstitials in disordered metallic alloys.

\begin{figure}[H]
    \centering
    \includegraphics[width=\linewidth]{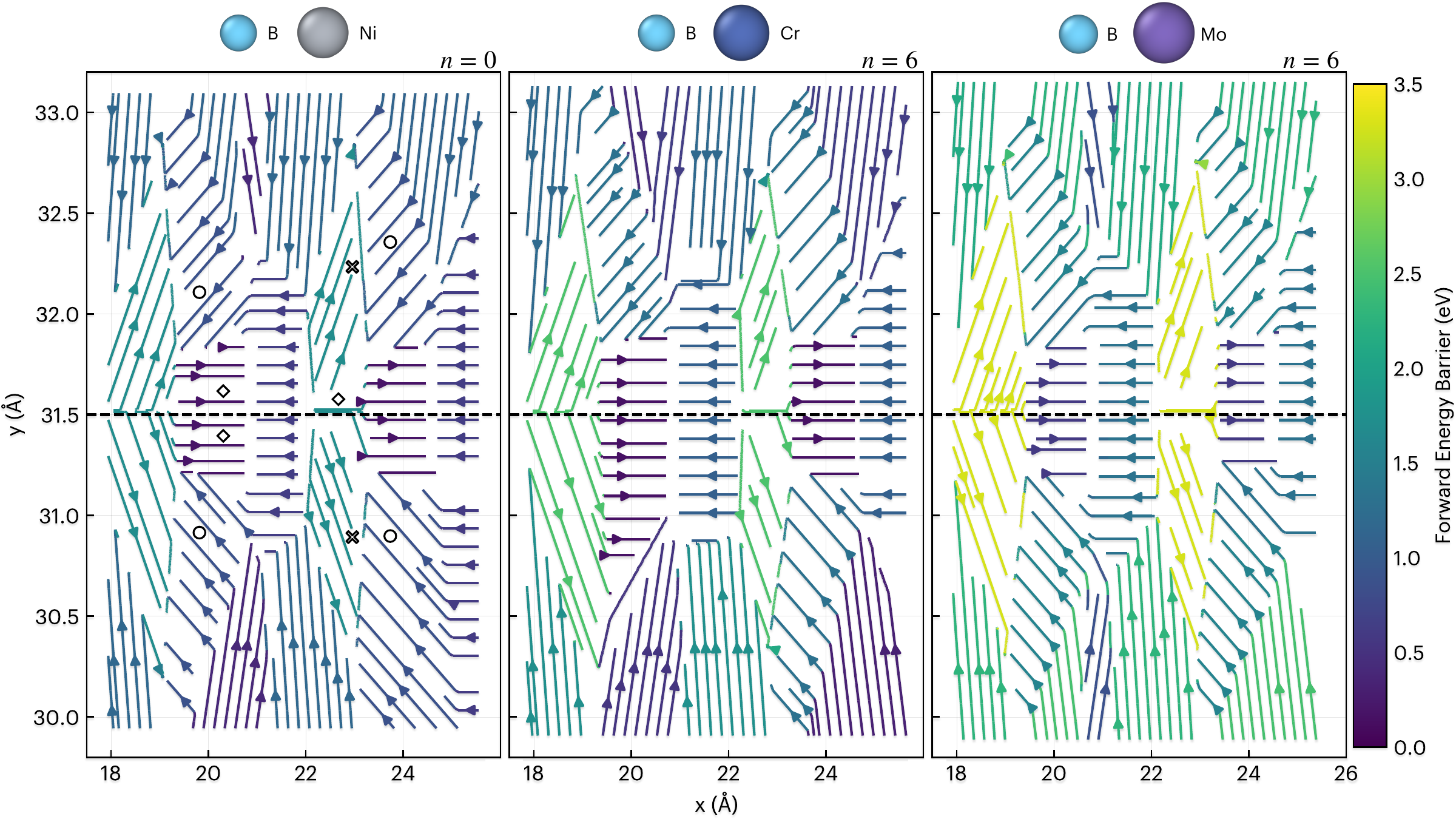}
    \caption{Displacement vector map of forward energy barriers for boron diffusion events at the $\Sigma5\,\langle100\rangle\,\{210\}$ (S5) grain boundary (GB), compiled from the same dataset used to generate the spectra in Fig.~\ref{fig:gb_spectra}. The GB midplane is indicated by the dashed black line. From left to right, the panels show boron diffusion in pure Ni ($n=0$), with six Cr atoms decorating the gaining cage ($n=6$), and with six Mo atoms decorating the gaining cage ($n=6$). Directional markers are overlaid to aid interpretation: diamonds represent in-plane diffusion along the GB midplane, circles denote diffusion toward the midplane, and ``x'' markers indicate diffusion away from the midplane.}
    \label{fig:see_diffusion}
\end{figure}

\begin{figure}[H]
    \centering
    \includegraphics[width=\linewidth]{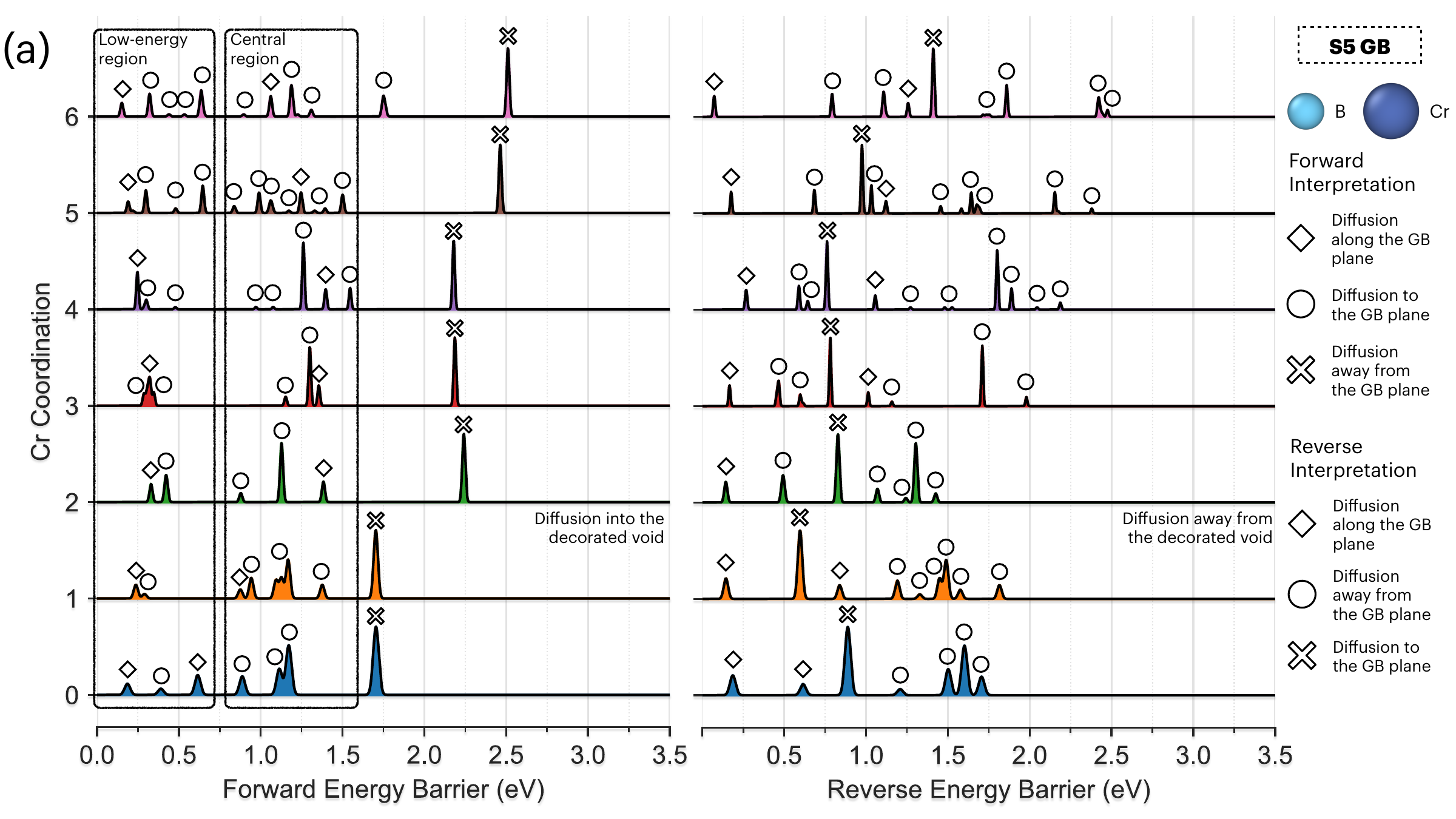}\\
    \includegraphics[width=\linewidth]{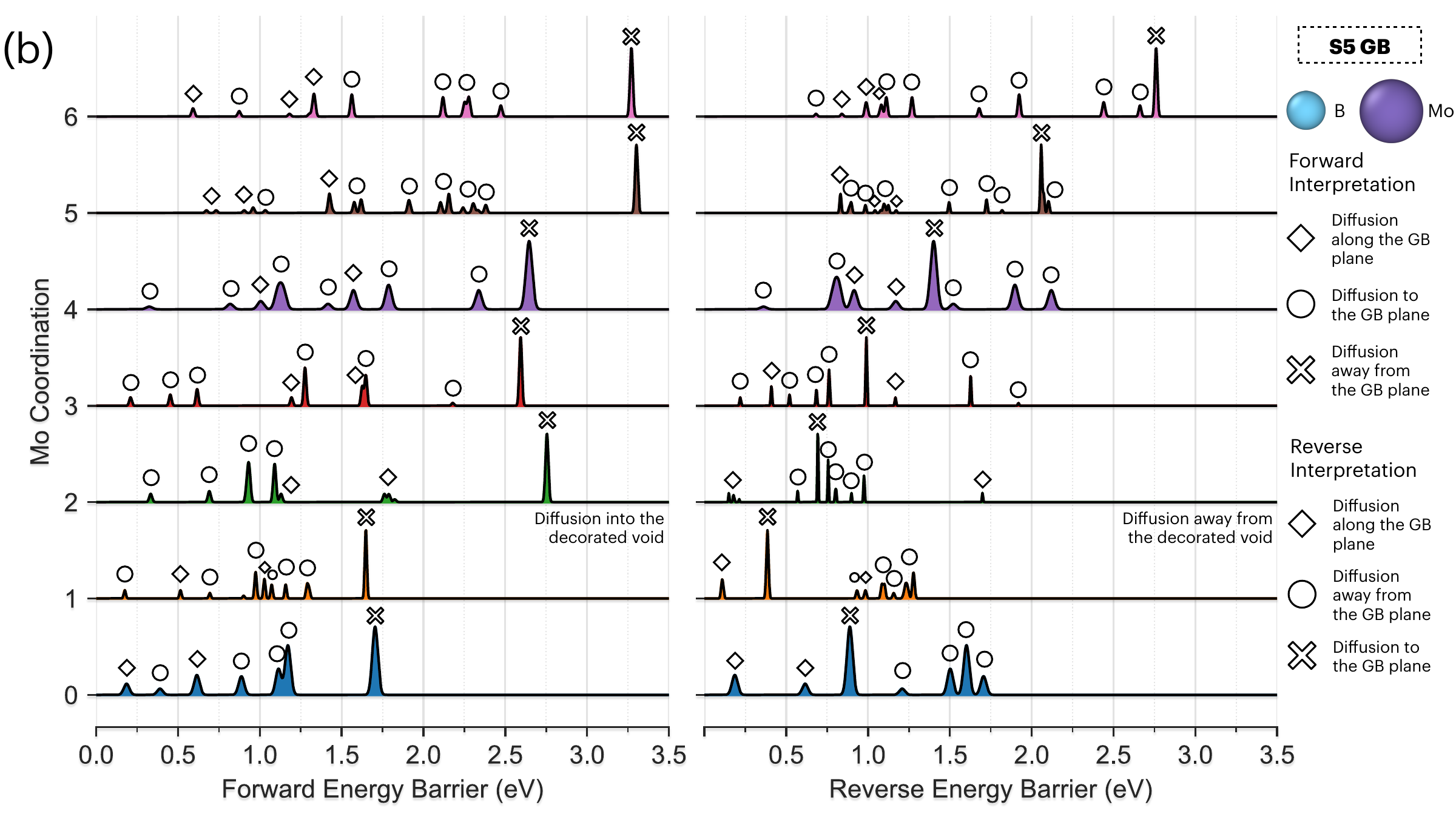}
    \caption{Activation energy spectra for boron diffusion in the $\Sigma5\,\langle100\rangle\,\{210\}$ (S5) grain boundary (GB). (a) Forward and reverse activation energy barriers as a function of Cr coordination in boron’s nearest-neighbor shell. (b) Same, as a function of Mo coordination. Colors indicate the number of Cr or Mo atoms surrounding the final-state GB site. Boxed regions correspond to the GB sites labeled in Fig.~\ref{fig:diff}a. Symbols denote the type of diffusion event underlying each peak; for example, diamonds represent hops between two voids located on the GB midplane, while ``x'' markers indicate diffusion away from the GB midplane. Note the inverse relationship between forward and reverse energy barrier spectra when interpreting the data.}
    \label{fig:gb_spectra}
\end{figure}

Before presenting the full analysis of boron migration at the S5 GB, it is instructive to visualize the diffusion events using the displacement vector map shown in Fig.~\ref{fig:see_diffusion}. As one moves from left to right, beginning with boron diffusion in pure Ni, then in environments with six nearest-neighbor Cr atoms, and finally with six nearest-neighbor Mo atoms, the color of the vectors progressively shifts toward warmer tones. This gradient provides a clear visual indication that Mo significantly suppresses boron mobility in all directions. The highest energy barriers correspond to diffusion events away from the GB midplane, while low-energy diffusion is observed along the GB midplane. Vectors pointing toward the GB midplane correspond to circle-marked barriers in Fig.~\ref{fig:gb_spectra}, those parallel to the GB midplane correspond to diamond-marked barriers, and those pointing away from the GB midplane correspond to ``x''-marked barriers. For clarity, the B--Ni vector map corresponds to the $n = 0$ forward energy barrier spectra shown in Fig.~\ref{fig:gb_spectra}a,b (blue curves). The B--Cr vector map corresponds to the $n = 6$ spectrum in Fig.~\ref{fig:gb_spectra}a (pink curve), and the B--Mo vector map corresponds to the $n = 6$ spectrum in Fig.~\ref{fig:gb_spectra}b (pink curve).

The boron migration energy barrier spectra for the S5 GB structure are compiled in Fig.~\ref{fig:gb_spectra}, with panel (a) presenting the B--Cr results and panel (b) showing the B--Mo results. Forward and reverse barriers are plotted side-by-side for each case to highlight directional asymmetry. In these spectra, peak amplitudes reflect the frequency with which distinct local motifs contribute to a given barrier height. A tall peak therefore indicates that many sampled configurations share the same chemical and structural environment, whereas smaller peaks arise from less common motifs. Peak height should thus be interpreted as motif multiplicity rather than an intrinsic probability or transition rate. Chemically complex environments, such as those with $n=5$ or $n=6$ solute neighbors, naturally give rise to a larger number of distinct local motifs, which appear as multiple smaller peaks distributed across the spectrum. As a reminder, forward barriers correspond to the energetic cost of entering the decorated void, while reverse barriers indicate the cost of leaving it. 

Owing to the compounded effects of structural disorder and chemical complexity, these spectra exhibit substantially richer features than their bulk counterparts. One consistent observation is that, relative to the bulk distributions (Fig.~\ref{fig:bulk_spectra}), boron migration within the GB region involves substantially lower energy barriers for in-plane diffusion (diamond markers). This trend reflects the enhanced mobility associated with planar defects, where disrupted atomic packing facilitates interstitial transport along the boundary. Such behavior aligns with the physical expectation that planar defects act as fast diffusion channels (diffusivity profiles provided in Fig.~\ref{fig:diff}b). 

The correlation between in-plane migration and lower energy barriers holds particularly well for a distinct set of stable interstitial sites near the GB midplane. These sites yield the lowest $E_{\mathrm{forward}}$ values, typically at or below 0.5~eV, and are labeled as the ``low-energy region'' in Fig.~\ref{fig:gb_spectra}a (diamond markers). A second group of in-plane sites, located within the box labeled ``central region,'' exhibits slightly higher forward barriers in the range of 1.0--1.5~eV (also indicated by diamond markers). By contrast, the upper end of the spectra is dominated by out-of-plane migration events, which involve transitions to sites located further from the GB midplane. These events are labeled with ``x'' markers in Fig.~\ref{fig:gb_spectra}. Representative configurations for each of these voids are provided in Fig.~S5 of the Supplemental Materials. Overall, the results reinforce the notion that in-plane voids, particularly those aligned with well-defined GB channels, serve as kinetically favorable pathways for interstitial diffusion. The peaks of the spectra are compiled in Table~\ref{tab:gb_barrier_spectra_peaks}.

\begin{table}[H]
\centering
\begin{tabular}{c|p{0.45\linewidth}|p{0.45\linewidth}}
\toprule\toprule
$\boldsymbol{n}$ & $\boldsymbol{E_{\mathrm{forward}}}$ \textbf{ (eV)} & $\boldsymbol{E_{\mathrm{reverse}}}$ \textbf{ (eV)} \\
\midrule
\multicolumn{3}{c}{\textbf{Cr (S5 GB)}} \\
\midrule
0 & 0.187, 0.398, 0.61,  0.88,  1.111, 1.169, 1.708 & 0.186, 0.61,  0.88,  1.207, 1.496, 1.592, 1.708 \\
1 & 0.241, 0.298, 0.867, 0.943, 1.095, 1.133, 1.171, 1.38,  1.703 & 0.147, 0.595, 0.848, 1.198, 1.334, 1.451, 1.49,  1.568, 1.821 \\
2 & 0.337, 0.429, 0.873, 1.132, 1.391, 2.242, & 0.151, 0.501, 0.831, 1.065, 1.24,  1.298, 1.434 \\
3 & 0.282, 0.321, 0.350, 1.154, 1.299, 1.357, 2.191 & 0.165, 0.469, 0.598, 0.617, 0.783, 1.013, 1.161, 1.714, 1.982 \\
4 & 0.243, 0.302, 0.479, 0.972, 1.080, 1.258, 1.395, 1.543, 2.183 & 0.272, 0.585, 0.644, 0.761, 1.054, 1.270, 1.485, 1.524, 1.798, 1.886, 2.042, 2.189 \\
5 & 0.187, 0.222, 0.303, 0.477, 0.650, 0.836, 0.986, 1.056, 1.172, 1.241, 1.334, 1.392, 1.496, 2.469 & 0.177, 0.679, 0.970, 1.037, 1.126, 1.450, 1.585, 1.640, 1.674, 2.154, 2.177, 2.378 \\
6 & 0.148, 0.316, 0.436, 0.532, 0.641, 0.893, 1.061, 1.193, 1.229, 1.313, 1.746, 2.515 & 0.073, 0.792, 1.109, 1.256, 1.414, 1.719, 1.755, 1.865, 2.426, 2.475 \\
\midrule
\multicolumn{3}{c}{\textbf{Mo (S5 GB)}} \\
\midrule
0 & 0.187, 0.398, 0.61,  0.88,  1.111, 1.169, 1.708 & 0.186, 0.61,  0.88,  1.207, 1.496, 1.592, 1.708 \\
1 & 0.513, 0.686, 0.898, 0.975, 1.033, 1.072, 1.149, 1.284, 1.65 & 0.394, 0.923, 0.982, 1.1, 1.158, 1.237, 1.276 \\
2 & 0.329, 0.686, 0.933, 1.094, 1.131, 1.760, 1.785, 1.822, 2.759 & 0.152, 0.175, 0.214, 0.567, 0.693, 0.756, 0.803, 0.897, 0.975, 1.697 \\
3 & 0.207, 0.450, 0.620, 1.190, 1.275, 1.627, 1.651, 2.173, 2.598 & 0.220, 0.409, 0.521, 0.685, 0.763, 0.987, 1.168, 1.625, 1.918 \\
4 & 0.318, 0.82,  1.006, 1.136, 1.415, 1.564, 1.787, 2.345, 2.643 & 0.355, 0.815, 0.907, 1.165, 1.404, 1.515, 1.901, 2.122 \\
5 & 0.671, 0.738, 0.898, 0.965, 1.032, 1.420, 1.580, 1.620, 1.915, 2.102, 2.155, 2.235, 2.302, 2.383, 3.305 & 0.834, 0.899, 0.984, 1.042, 1.075, 1.094, 1.127, 1.172, 1.497, 1.725, 1.822, 2.056, 2.076, 2.102 \\
6 & 0.589, 0.875, 1.175, 1.298, 1.325, 1.557, 2.116, 2.252, 2.279, 2.470, 3.274 & 0.678, 0.837, 0.985, 1.081, 1.113, 1.271, 1.684, 1.928, 2.437, 2.659, 2.765 \\
\bottomrule
\end{tabular}
\caption{Peak locations in the forward and reverse energy spectra shown in Fig.~\ref{fig:gb_spectra}. The $n=0$ datasets represent undoped configurations with no Cr or Mo atoms within the boron first-nearest neighbor shell.}
\label{tab:gb_barrier_spectra_peaks}
\end{table}

Having established the general trends in Fig.~\ref{fig:gb_spectra}, now begins a more detailed examination of the B--Cr spectra presented in Fig.~\ref{fig:gb_spectra}a. Starting with a chemically homogeneous GB environment ($n = 0$, blue curve), the kinetic landscape is broadened into an almost continuous low-energy distribution, with notable peaks near 1.169 and 1.708~eV. Within this spectrum, the in-plane diffusion events (diamond markers) occur at a much lower barrier, 0.187~eV, compared to the bulk counterpart of 1.484~eV, highlighting the efficiency of midplane pathways in the absence of solute decoration. Introducing a single Cr neighbor ($n = 1$, orange curve) sharpens the low-energy region into a prominent peak near 0.241~eV. This region shows a combination of in-plane diffusion events (diamonds, 0.241~eV) and boron transfer from out-of-plane voids into the GB midplane (circles, 0.298~eV), indicating that even minimal Cr decoration alters the partitioning between in-plane and out-of-plane migration. As Cr coordination increases to $n = 2$ (green curves), the forward barrier spectrum for out-of-plane migration (``x'' markers) exhibits a sharp rise, continuing to escalate through $n = 4$ and reaching 2.515~eV at $n = 6$. This terminal value substantially exceeds the bulk counterpart of 1.761~eV for the same Cr coordination, signifying that Cr effectively suppresses boron escape from the GB midplane. By contrast, the low-energy in-plane diffusion barriers (diamond markers) remain relatively stable across all $n$, suggesting that the thermodynamically favored midplane voids retain kinetically accessible pathways even under heavy Cr decoration. Meanwhile, the low-barrier events marked by circles show a steady increase with $n$, indicating that Cr progressively impedes boron from accessing midplane voids from out-of-plane sites.

In contrast to the bulk behavior, the $E_{\mathrm{reverse}}$ spectrum for B--Cr at the S5 GB exhibits a more nuanced and spatially dependent evolution. A significant fraction of reverse barriers shift downward with increasing Cr coordination, but unlike in the bulk, this shift is non-uniform across the GB region. The circle-marked barriers shift upward, indicating that diffusion from void sites along the GB midplane back to out-of-plane voids becomes more costly, which is consistent with the thermodynamic preference for midplane occupancy. By contrast, the diamond-marked barriers shift slightly downward, suggesting, as in the bulk, a modest bias for boron to migrate away from the Cr-decorated void more easily than towards it, although the effect is far less pronounced at the GB. Out-of-plane migration events (``x'' markers) show a substantial drop in reverse barrier energy. Under the reverse-barrier interpretation, this corresponds to a reduced cost for boron to transition away from decorated out-of-plane voids back into the GB midplane, where midplane sites are energetically preferred. Consistently, migration into out-of-plane positions is associated with high forward barriers (``x'' markers in the $E_{\mathrm{forward}}$ spectra) and high reverse penalties (circle markers in the $E_{\mathrm{reverse}}$ spectra), reinforcing the chemical and geometric disfavorability of boron escape. These results suggest that Cr-rich GBs act as effective sinks for boron: they support relatively fast in-plane diffusion while imposing steep out-of-plane barriers that suppress migration into the surrounding grains. This directional asymmetry underscores the dual role of Cr as a kinetic guide and a moderate thermodynamic stabilizer, with implications for GB-mediated transport, dopant confinement, and long-term boron retention in chemically complex alloys.

The energy barrier landscape for B--Mo diffusion (Fig.~\ref{fig:gb_spectra}b) differs markedly from that of B--Cr, largely due to the stronger local distortion imposed by Mo atoms. This distinction is most apparent in the fact that circle-marked diffusion events have lower barriers than diamond-marked events. Unlike Cr, which promotes diffusion along the GB midplane, Mo reduces boron mobility along the GB midplane. Instead, boron more readily accesses low-energy void cages via diffusion from out-of-plane positions. This provides clear evidence that Mo presents a substantial steric hindrance to boron motion along the GB midplane. Interestingly, once $n = 5$ or $n = 6$, the trend reverses and diamond-marked events again correspond to the lowest barriers. This indicates that heavily Mo-coordinated void cages along the GB midplane are most accessible through diffusion confined to the GB midplane itself. The distinction between the ``low-energy region'' and the ``central region'' is also less apparent in the case of Mo. This arises because Mo uniformly raises the forward migration barriers at the GB, shifting a large fraction of events into the 0.5~eV to 2.0~eV range for both low-energy and central void sites.

As with Cr, the highest forward energy barriers in Fig.~\ref{fig:gb_spectra}b correspond exclusively to out-of-plane migration events, but in the case of Mo these values are larger by approximately 0.76~eV. This confirms that Mo surpasses Cr in its ability to trap boron at the GB. In addition, unlike Cr, Mo raises the forward energy barrier for boron to enter the thermodynamically preferred midplane void cages, with the lowest diamond-marked barrier reaching 0.589~eV at $n = 6$ compared to only 0.148~eV for $n = 6$ Cr. Taken together, these results show that Mo exerts a dual influence on boron transport. It increases the barriers for in-plane diffusion into stable midplane sites while simultaneously imposing even steeper penalties on out-of-plane migration. The combined effect is that Mo not only confines boron more strongly to the GB region, but also suppresses its overall mobility within the midplane, reinforcing Mo’s role as both a kinetic and thermodynamic trap for interstitial boron.

Similar to the B--Cr spectra, the B--Mo reverse spectrum exhibits a variable shift relative to its corresponding forward barrier distribution. In the case of B--Mo, the diamond-marked barriers shift substantially toward lower energies, indicating a strong kinetic bias that makes it easier for boron to diffuse back into undecorated midplane voids. A similar trend is observed for out-of-plane migration events (``x'' markers), where the reverse process back to the midplane is energetically favored relative to the forward event. However, the $n=6$ Mo off-plane void is an interesting example (``x'' marked reverse event in the pink spectrum of Fig.~\ref{fig:gb_spectra}b), as it exhibits the highest reverse migration barrier within the $n=6$ Mo spectrum. This is starkly different from pure Ni or $n=6$ Cr. This penalty arises because the symmetric, bulk-like coordination of this off-plane void site, combined with the large atomic radius of Mo, creates strong steric hindrance. Meaning that specific off-plane void is very difficult to both get into (high forward ``x'' barrier) and get out of (high reverse ``x'' barrier) when it has 6 Mo atoms making up the cage. The local trapping mechanism at this off-plane void is so dominant that the reverse barrier becomes the highest barrier observed. This agrees well with the bulk diffusion findings for Mo, where it proved a strong trap against reverse boron migration events while Cr and pure Ni did not. Moving beyond this point, the circle-marked diffusion events shift upward in the reverse spectrum, indicating that diffusion away from midplane voids becomes increasingly difficult with higher Mo decoration. At $n = 6$, the high-barrier circle-marked peaks for both B--Mo and B--Cr confirm that these midplane voids act as effective traps. However, the B--Mo barrier is approximately 0.5~eV higher than that of B--Cr, underscoring that Mo is the more proficient trapping agent for boron at the GB. 

Overall, these spectra reveal that the migration landscape is not only directionally anisotropic but also highly sensitive to chemical decoration of the GB void cages. In the chemically homogeneous case ($n = 0$, blue curves in Fig.~\ref{fig:gb_spectra}), the spectrum is comparatively simple: two diamond-marked peaks, four circle-marked peaks, and a single ``x''-marked peak. This limited set of features indicates that geometric disorder alone produces only a few kinetically distinct migration pathways. With the introduction of solutes, however, the number of distinct peaks increases dramatically. At $n = 6$, the spectrum still retains the two diamond-marked peaks and the single ``x''-marked peak observed at $n = 0$, showing that the original midplane and out-of-plane pathways remain accessible. In addition, the spectrum now contains nine separate circle-marked peaks for Cr and six for Mo, reflecting new nonequivalent events for migration into the chemically decorated GB void cages. This proliferation of discrete features demonstrates that chemical heterogeneity introduces a richer set of migration pathways than geometry alone. Moreover, chemical decoration enhances forward--reverse asymmetry, where barriers for leaving chemically decorated sites often differ substantially from those for entering them. In this way, the transition landscape evolves from only structurally defined at $n = 0$ to complex, chemically discriminated, and directionally biased at higher $n$. These findings highlight the importance of solute identity and placement in shaping interstitial mobility, and motivate a closer examination of segregation energies as the thermodynamic counterpart to the kinetic constraints revealed in the migration spectra.

\subsection*{Diffusivity and Macroscopic Transport Implications}

\begin{figure}[H]
    \centering
    \includegraphics[width=\linewidth]{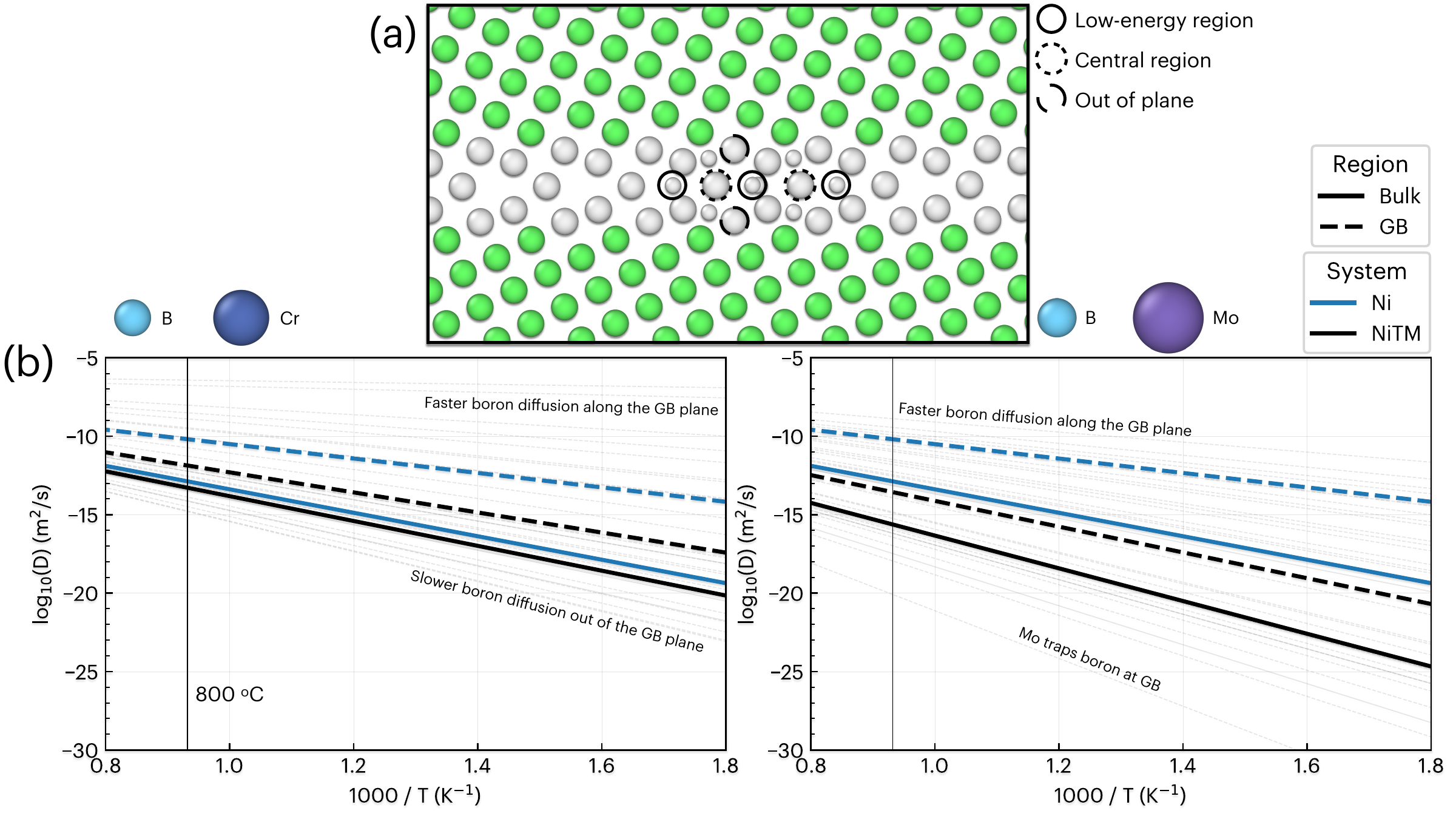}
    \caption{(a) Visualization of the void network (smaller atoms) within the $\Sigma5\,\langle100\rangle\,\{210\}$ grain boundary (S5 GB), with key spectral regions labeled. Out-of-plane voids are unlabeled unless situated directly beneath visible Ni lattice atoms. (b) Macroscopic diffusivity profiles for boron in Ni--Cr and Ni--Mo systems, computed using Eqs.~\ref{eq:diff} and~\ref{eq:eeff}. Bold black lines represent the average boron diffusivity across $n = 1$ to $6$ coordination levels in transition metal (TM)--decorated bulk (solid) and S5 GB (dashed) environments. Bold blue lines show boron diffusivity in pure Ni. Semi-transparent curves trace individual diffusivity trajectories derived from the full activation energy spectrum. A vertical line marks $T = 800\,^\circ$C.}
    \label{fig:diff}
\end{figure}

To translate the site-specific migration barriers identified in the spectra into macroscopic transport properties, we employ an Arrhenius-type relation \cite{mehrerDiffusionSolids2007}:

\begin{equation}\label{eq:diff}
    D = D_0 \exp\left( -\frac{E_a}{k_B T} \right)
\end{equation}
where $ D_0 = \nu a^2 $ is the pre-exponential factor \cite{mehrerDiffusionSolids2007}, $ E_a $ is the activation energy (eV), $ k_B $ is the Boltzmann constant, $ T $ is the temperature, $ \nu \sim 10^{13}\,\text{s}^{-1} $ is the attempt frequency, and $ a = 3.52\,\text{\AA} $ is the lattice parameter. This yields $ D_0 = 1.24 \times 10^{-6}\,\text{m}^2/\text{s} $, consistent with typical interstitial diffusion prefactors in metallic systems \cite{wangDiffusionBoronAlloys1995}. Using the forward ($E_{\mathrm{forward}}$) and reverse ($E_{\mathrm{reverse}}$) migration barriers, an effective activation energy was defined as

\begin{equation}\label{eq:eeff}
    E_{\mathrm{a}} = \frac{\langle E_{\mathrm{forward}} \rangle + \langle E_{\mathrm{reverse}} \rangle}{2},
\end{equation}
where the angle brackets denote an average over $n = 5,6$ for Cr- or Mo-coordinated environments. This formulation approximates the collective aggregation of diffusional behavior that would be observed experimentally. The resulting effective barriers were then used in conjunction with Eq.~\ref{eq:diff} to construct macroscopic diffusivity profiles for boron in the bulk (solid black lines in Fig.~\ref{fig:diff}b) and at the S5 GB (dashed black lines). The semi-transparent lines trace the ensemble of individual B--Cr and B--Mo diffusivity curves, each derived from a distinct effective activation energy, which were averaged to yield the bold macroscopic trends.

Macroscopically, Fig.~\ref{fig:diff}b shows that Mo reduces boron diffusivity by approximately two orders of magnitude compared to Cr, both in the bulk and along the S5 GB at 800~$^\circ$C. Relative to pure Ni (blue lines), Mo suppresses boron mobility more severely than Cr. At the atomistic level, the semi-transparent lines reveal that boron retains access to ``high-speed'' transport pathways in low-barrier regions to and along the GB midplane for Cr ($D \sim 10^{-6}$~m\textsuperscript{2}/s), but less so for Mo ($D \sim 10^{-8}$~m\textsuperscript{2}/s). While both Cr and Mo inhibit boron egress from the S5 GB, Mo is significantly more effective at kinetically trapping boron (e.g., $D_{\mathrm{GB}}\sim10^{-20}$~m\textsuperscript{2}/s at 800 $^\circ$C) relative to Cr (e.g., $D_{\mathrm{GB}}\sim10^{-15}$~m\textsuperscript{2}/s at 800 $^\circ$C), a trend that also holds in the bulk. Additionally, the wide spread of individual diffusivity trajectories underscores how substitutional solutes create highly heterogeneous diffusion environments for mobile interstitials capable of \textit{direct interstitial migration}. These computational findings are consistent with experimental observations by Tytko et al.~\cite{tytkoMicrostructuralEvolutionNibased2012}, who reported co-segregation of boron and Mo at GBs and at the interfaces of Cr- and Mo-rich M\textsubscript{23}C\textsubscript{6} carbides, and further identified boron substituting for carbon within Mo-rich M\textsubscript{23}C\textsubscript{6}. Complementing these results, Kontis et al.~\cite{kontisEffectBoronGrain2016a, kontisRoleBoronImproving2017a} demonstrated that boron preferentially partitions into Cr-rich M\textsubscript{5}B\textsubscript{3} borides along GBs, serving as the dominant boron reservoir with negligible amounts in solid solution. A deeper analysis of trapping behavior is provided in the following section on the segregation energy spectra.

\subsection*{Segregation Energy Spectra}

\begin{figure}[H]
    \centering
    \includegraphics[width=\linewidth]{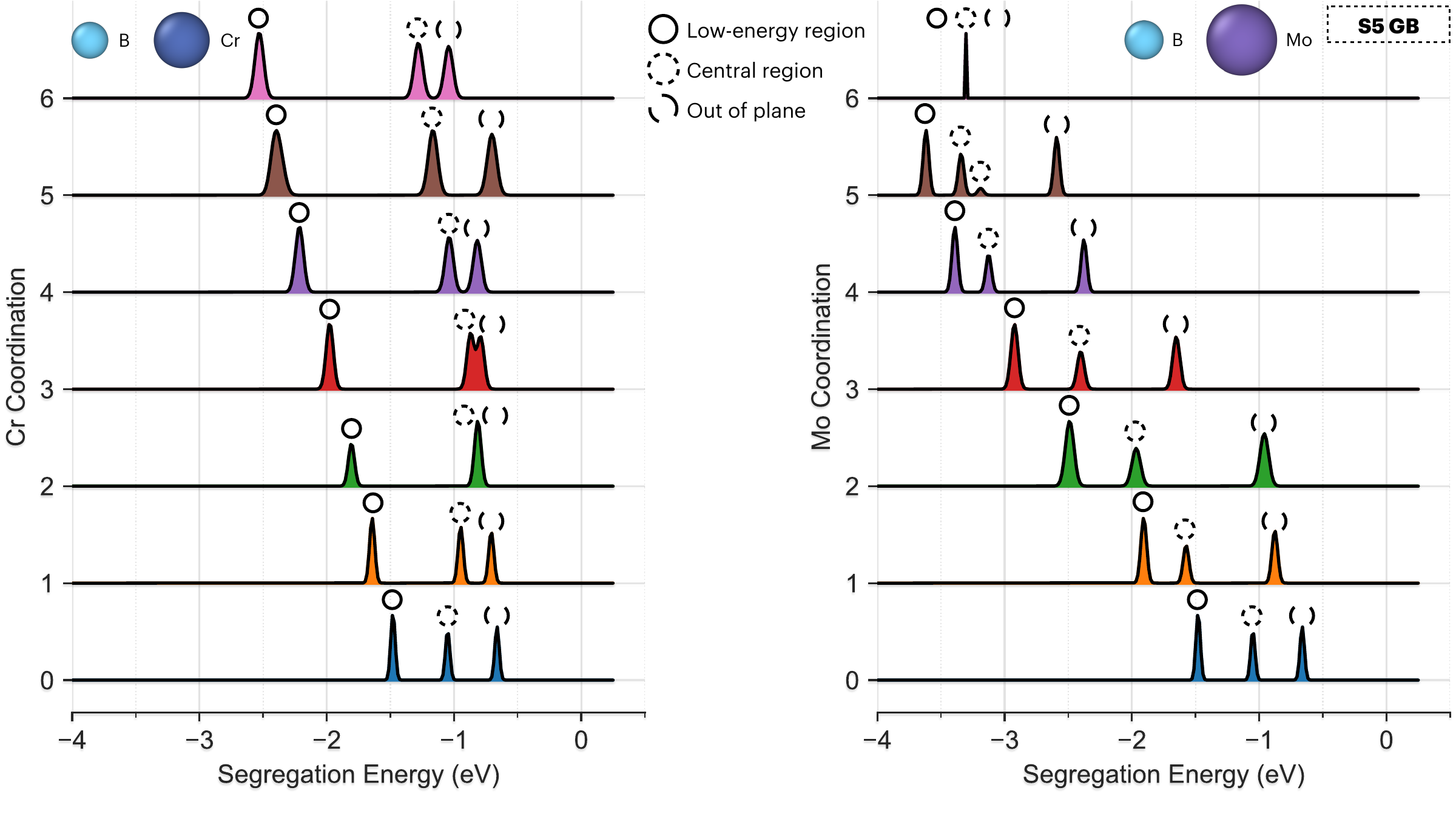}
    \caption{Segregation energy spectra for boron at the $\Sigma5\,\langle100\rangle\,\{210\}$ (S5) grain boundary (GB) as a function of Cr or Mo coordination. Color corresponds to the number of solute atoms coordinated to the boron site.}
    \label{fig:eseg_spectra}
\end{figure}

The segregation energy spectra are presented in Fig.~\ref{fig:eseg_spectra}, which provide complementary thermodynamic insights into boron--solute interactions at the GB. In the B--Cr system, a clear trend is observed: as Cr atoms are incrementally introduced around the boron site (progressing from blue to pink curves), the segregation energy $E_{\mathrm{seg}}$ becomes increasingly negative. This indicates that replacing B--Ni bonds with B--Cr bonds has a net stabilizing effect on boron incorporation at the GB, regardless of the specific site. The most negative $E_{\mathrm{seg}}$ values correspond to thermodynamically preferred voids along the GB midplane. These are the same sites that exhibited low forward diffusion barriers and elevated reverse barriers, promoting boron retention. In contrast, the intermediate values reflect less stable midplane voids, while the least negative $E_{\mathrm{seg}}$ values correspond to sites out of the GB midplane.

In the B--Mo system, $E_{\mathrm{seg}}$ stabilization is more pronounced than in the B--Cr case. As with Cr, the most negative $E_{\mathrm{seg}}$ values correspond to GB midplane voids. A clear ``downward'' trend from blue to pink curves reflects the systematic stabilization of boron with increasing Mo coordination. A particularly notable feature emerges at $n = 6$: unlike the broader energy distributions observed at lower coordination numbers, the $E_{\mathrm{seg}}$ spectrum at full Mo coordination exhibits a sharp, delta-like peak. This suggests that once boron is fully coordinated by Mo in the S5 GB region, its thermodynamic stabilization becomes largely independent of the specific void it occupies. Supporting this interpretation, Fig.~S4b shows that boron samples a wide range of spatially distinct $n = 6$ B--Mo configurations, all yielding nearly identical stabilization energies.  

While both Cr and Mo promote directional anisotropy in boron diffusion at the GB, their trapping strengths diverge. As shown in the activation energy spectra and diffusivity profiles (Fig.~\ref{fig:gb_spectra} and Fig.~\ref{fig:diff}b), Cr facilitates fast in-plane migration while presenting high barriers to out-of-plane diffusion, whereas Mo produces an overall reduction in boron mobility. The $E_{\mathrm{reverse}}$ spectra further confirm that boron is consistently ``pushed'' back toward the GB midplane, with escape from out-of-plane sites becoming increasingly unfavorable as solute coordination rises.

These findings indicate complementary roles for Cr and Mo in governing boride formation through the coupling of segregation thermodynamics and interfacial transport kinetics. Cr-rich GB sites act as kinetic guides, facilitating rapid in-plane boron migration along the boundary and discouraging escape into the surrounding matrix. At the same time, Cr provides a favorable thermodynamic environment for boron segregation, consistent with the frequent observation of Cr-rich borides in Ni alloys. Mo-rich sites build on this effect by not only suppressing redistribution through higher migration barriers, but also acting as particularly strong thermodynamic anchors: their segregation wells are consistently deeper and more site-insensitive than those of Cr. As a result, both Cr and Mo promote boron enrichment at GB sites, with Cr enhancing delivery and local stabilization and Mo reinforcing long-term retention. This accumulation of boron, as demonstrated in our previous work~\cite{dolezalSegregationOrderingLight2025}, lowers the GB energy $\gamma_{\mathrm{GB}}$, thereby reducing the thermodynamic driving force for boundary migration.

In classical nucleation theory~\cite{porterPhaseTransformationsMetals2021}, increasingly negative $E_{\mathrm{seg}}$ for B--Cr, and especially for B--Mo, combined with their kinetic trapping tendencies, enhances the local supersaturation, $S$, of boron and its preferred metallic partners. This increases the magnitude of the volumetric free energy change, $\lvert \Delta G_V \rvert$, thereby reducing both the critical radius,  
\begin{equation}
r^* = \frac{-2\gamma}{\Delta G_V},
\end{equation}
and the nucleation barrier,  
\begin{equation}
\Delta G^* = \frac{16\pi\gamma^3}{3(\Delta G_V)^2}S(\theta),
\end{equation}
where $\gamma$ is the interfacial energy between the boride precipitate and matrix, and $\Delta G_V$ is the volumetric free energy change driving nucleation, and $S(\theta)$ is a geometric factor accounting for heterogeneous nucleation, which depends on the wetting angle $\theta$. Because boron is highly mobile along clean and Cr-decorated GB midplanes, it can rapidly sample available interstitial sites. Trapping therefore depends on the slower vacancy-mediated diffusion of substitutional solutes into the GB environment. Faster Cr diffusion~\cite{murarkaDiffusionChromiumNickel1964} enables efficient delivery of traps, whereas Mo diffuses more slowly~\cite{swalinDiffusionMagnesiumSilicon1957, divyaInterdiffusionNiMo2010} but generates deeper thermodynamic wells that strongly retain boron once encountered. Together, these effects favor nucleation of stable borides such as Cr\textsubscript{5}B\textsubscript{3}, Cr\textsubscript{2}B, and Cr(BMo)\textsubscript{2}~\cite{kontisEffectBoronGrain2016a, duPrecipitationEvolutionGrain2017a, tekogluMetalMatrixComposite2024b, tekogluSuperiorHightemperatureMechanical2024c, seoEtaPhaseBoride2007a, guptaCompositionallyGradedNanosized2021, gaoEffectTiB2Content2023, kitaguchiMCDecompositionBoride2024, baeAdditiveManufacturingStrong2025a}. Importantly, the competition between Cr-only borides and mixed Cr/Mo borides is highly sensitive to the Mo concentration in the Ni-based superalloy matrix: low-Mo alloys stabilize Cr-rich phases such as Cr\textsubscript{5}B\textsubscript{3}, whereas higher Mo contents favor mixed Cr(BMo)\textsubscript{2} structures through enhanced B--Mo co-segregation. When the boron content is below the supersaturation threshold for nucleation, elemental boron instead co-segregates at Cr- and/or Mo-rich M\textsubscript{23}C\textsubscript{6} $\gamma$/carbide phase boundaries~\cite{tytkoMicrostructuralEvolutionNibased2012}. 

Once formed, GB-anchored borides impede high-temperature redistribution by reducing the effective vacancy flux along the boundary. As given in Fig.~\ref{fig:eseg_spectra}a, Cr- and Mo-rich environments stabilize B at the GB ($E_{\mathrm{seg}} < 0$). These motifs reduce B mobility, with Mo producing system-wide slowdowns and both Cr and Mo strongly suppressing out-of-plane loss (Fig.~\ref{fig:diff}b). We take these trends as proxies for a stiffer, more bonded GB network that very likely raises vacancy formation and migration barriers for GB self-diffusion, thereby lowering the effective $D_{\mathrm{GB}}$ relative to pure Ni. According to the Coble creep relation~\cite{kassnerFundamentalsCreepMetals2009},
\begin{equation}\label{eq:coble}
\dot{\varepsilon} \propto \frac{D_{\mathrm{GB}} \, \sigma \, \Omega}{k_B T \, d^3},
\end{equation}
a reduction in $D_{\mathrm{GB}}$ directly suppresses the vacancy-mediated grain-boundary sliding that drives Coble creep. Holding $\sigma$, $T$, and $d$ fixed, a one to two order-of-magnitude decrease in $D_{\mathrm{GB}}$ translates to a commensurate reduction in $\dot{\varepsilon}$. In addition, the presence of stable boride particles at GBs exerts a Zener pinning force, expressed as a pinning pressure~\cite{smith1948introduction}, 

\begin{equation} 
P_Z = \frac{3f\gamma}{2r}, 
\end{equation}
where $f$ is the particle volume fraction and $r$ is the particle radius, further reducing GB mobility and contributing to grain refinement. This synergy between Cr-mediated delivery of boron and Mo-driven thermodynamic retention produces a GB environment that is both kinetically and thermodynamically optimized for boride nucleation, structural stabilization, and creep resistance in Ni-based superalloys~\cite{garosshenEffectsZrStructure1987,kontisEffectBoronGrain2016a, kontisRoleBoronImproving2017a, tekogluSuperiorHightemperatureMechanical2024c}.

While this study focuses on interstitial migration, the same interplay between thermodynamic traps and kinetic pathways identified for B--Cr and B--Mo also applies to substitutional solutes~\cite{xingNeuralNetworkKinetics2024, xingShortrangeOrderLocalizing2022}. In substitutional systems, additional factors such as larger atomic radii, bonding character, and solvent--solute interactions can further amplify migration heterogeneity. This heterogeneity is explicitly captured in the activation energy barrier spectra (Fig.~\ref{fig:gb_spectra}) and diffusivity profiles (Fig.~\ref{fig:diff}b). These reveal a wide spread in boron mobility even at fixed solute coordination levels, arising from local structural asymmetries and site-specific energetics. Recent studies by Wagih et al. demonstrated that substitutional solutes exhibit site-specific segregation energy spectra at GBs, underscoring the thermodynamic variability imposed by local structural and chemical environments~\cite{wagihSpectrumGrainBoundary2019, wagihLearningGrainBoundarySegregation2022}. 

These insights, combined with the present findings, support a broader reinterpretation of diffusion in chemically complex systems such as high-entropy alloys (HEAs). The concept of ``sluggish'' diffusion was originally proposed as a core effect, attributing reduced atomic mobility to severe chemical disorder~\cite{yehNanostructuredHighEntropyAlloys2004b, tongMicrostructureCharacterizationAlx2005, changInfluenceSubstrateBias2008, ngEntropydrivenPhaseStability2012}. This hypothesis has been examined through experimental tracer diffusion studies and averaged activation energy analyses, which often report suppressed diffusivities relative to conventional alloys~\cite{vermaDiffusionHighEntropy2024}. However, such interpretations can obscure the underlying heterogeneity of diffusion pathways. For instance, Chen et al. observed reduced interdiffusion rates for Al, Co, Cr, and Ti in an Al--Co--Cr--Fe--Ni--Ti HEA, yet reported enhanced diffusion for Fe, highlighting the element-specific nature of migration kinetics~\cite{chenDiffusionBehaviorsFace2019}. Similarly, D\k{a}browa et al. questioned the universality of the sluggish diffusion effect, showing that certain HEA systems exhibit normal or even accelerated diffusion depending on temperature and composition, emphasizing the decisive role of local chemical environments~\cite{dabrowaDemystifyingSluggishDiffusion2019}. Such observations further support the need for spectrum-based approaches, as employed here, to resolve the full distribution of migration barriers rather than relying on single averaged values.

The present spectral analysis captures this heterogeneity directly, showing that diffusivity in chemically complex and structurally asymmetric systems is not uniformly suppressed but instead governed by a rugged, site-specific potential energy landscape. Diffusion does not proceed through a single representative barrier, but rather across a distribution of local transition states shaped by solute identity and coordination asymmetry. This framework connects atomistic mechanisms to macroscopic transport and provides a transferable model for interpreting diffusion in disordered alloys, consistent with recent studies on vacancy-mediated migration~\cite{xingNeuralNetworkKinetics2024, xingShortrangeOrderLocalizing2022}. In substitutional systems, vacancy-mediated migration pathways are likewise highly sensitive to the surrounding chemical and structural environment, further broadening the spectrum of possible barriers. These results demonstrate that spectrum-based approaches are essential for accurately resolving diffusion kinetics in complex alloys, and they caution against oversimplification into a single averaged activation energy when interpreting migration mechanisms or designing new alloy chemistries.

\section*{Conclusion}

This study introduces a spectral framework for analyzing light interstitial diffusion in chemically complex Ni-based alloys, revealing how local Cr and Mo coordination shapes the activation energy landscape for boron migration. By systematically sampling atomic environments in both bulk FCC Ni and a $\Sigma5\,\langle100\rangle\,\{210\}$ grain boundary, we construct migration barrier spectra that capture the coupled influence of structural topology and chemical disorder. The results demonstrate that interstitial diffusion is not governed by a single representative energy barrier, but instead by a chemically and spatially heterogeneous spectrum of transition states.  

In the bulk, migration energy distributions for boron exhibit distinct modality tied to solute identity and spatial arrangement. Both Cr and Mo raise migration barriers in symmetric configurations but induce directional asymmetry in partially decorated cages. At the grain boundary, Cr preserves low-barrier in-plane mobility while raising out-of-plane barriers that confine boron to midplane voids, whereas Mo imposes an across-the-board reduction in boron mobility. Macroscopic diffusivity analysis shows that both Cr and Mo reduce boron transport in Ni. On average, Mo suppresses boron diffusivity by an additional two orders of magnitude compared to Cr at 800~$^\circ$C. Moreover, Mo confines boron to the grain boundary much more strongly than Cr, reducing out-of-plane boron diffusivity by roughly five orders of magnitude relative to Cr. 

Both Cr and Mo promote boron segregation to grain boundaries by producing negative segregation energies, thereby creating thermodynamically favorable environments for interstitial trapping. Cr facilitates rapid in-plane boron migration while simultaneously stabilizing boron at GB sites, consistent with the prevalence of Cr-rich borides reported in Ni alloys. Mo builds on this effect by providing deeper and more uniform segregation wells across all coordination levels, establishing itself as a particularly strong thermodynamic anchor. Together, these results establish that substitutional solutes fundamentally reshape interstitial point defect diffusion spectra, controlling mobility in both bulk Ni and along crystallographic defects. This highlights a general paradigm in which substitutional chemistry dictates point defect kinetics at multiple length scales.  

These findings explain experimental observations of elemental boron segregation to Mo-rich carbide phase boundaries, Cr-rich grain-boundary--anchored borides, and (Mo,Cr)-rich grain-boundary--anchored borides in Ni-based superalloys. They also highlight how elemental identity governs the balance between mobility and retention at interfaces. More broadly, the spectral approach developed here provides a transferable methodology for capturing migration heterogeneity in disordered systems, with natural extensions to substitutional diffusion and chemically complex alloys such as HEAs. By treating diffusion as a sampling problem over a rugged energy landscape, this work establishes a foundation for predictive modeling of transport phenomena and for designing alloys in which diffusion spectra, rather than average barriers alone, are engineered for high-performance applications.

\section*{Author Contributions}
\textbf{T.D.D} Writing - original draft, writing - review \& editing, visualization, validation, software, methodology, investigation, formal analysis, data curation. \textbf{R.F.} Project administration and supervision, writing - review \& editing. \textbf{J. Li} Project administration and supervision, writing - review \& editing, funding acquisition. All authors contributed to the conceptualization of this project.

\section*{Data Availability}
The spectral scanning routine, all data generated from this work, and post-processing scripts will be provided at \url{https://github.com/tylerdolezal}.

\section*{Acknowledgments}
J. Li acknowledges support from National Science Foundation, USA CMMI-1922206 and DMR-1923976.

\bibliographystyle{ieeetr}

\begin{thebibliography}{10}

\bibitem{mehrerDiffusionSolids2007}
H.~Mehrer, {\em Diffusion in {{Solids}}}, vol.~155 of {\em Springer {{Series}} in {{Solid-State Sciences}}}.
\newblock Berlin, Heidelberg: Springer, 2007.

\bibitem{simsSuperalloysIIHighTemperature1987}
C.~T. Sims, N.~S. Stoloff, and W.~C. Hagel, eds., {\em Superalloys {{II}}: {{High-Temperature Materials}} for {{Aerospace}} and {{Industrial Power}}}.
\newblock New York: Wiley-Interscience, 2nd edition~ed., Sept. 1987.

\bibitem{bashirEffectInterstitialContent1993a}
S.~Bashir and M.~C. Thomas, ``Effect of interstitial content on high- temperature fatigue crack propagation and low- cycle fatigue of alloy 720,'' {\em Journal of Materials Engineering and Performance}, vol.~2, pp.~545--550, Aug. 1993.

\bibitem{tianSynergisticEffectsBoron2024b}
Q.~Tian, S.~Huang, H.~Qin, R.~Duan, C.~Wang, and X.~Lian, ``Synergistic {{Effects}} of {{Boron}} and {{Rare Earth Elements}} on the {{Microstructure}} and {{Stress Rupture Properties}} in a {{Ni-Based Superalloy}},'' {\em Materials}, vol.~17, p.~2007, Jan. 2024.

\bibitem{gongMicrostructuralEvolutionMechanical2023a}
X.~Gong, Y.~Wu, Z.~Gao, Y.~Sun, Y.~Guan, X.~Guan, X.~Qin, J.~Hou, and L.~Zhou, ``Microstructural {{Evolution}} and {{Mechanical Properties}} of a {{Ni-Based Alloy}} with {{High Boron Content}} for the {{Pre-Sintered Preform}} ({{PSP}}) {{Application}},'' {\em Materials}, vol.~16, p.~7483, Jan. 2023.

\bibitem{garosshenEffectsZrStructure1987}
T.~J. Garosshen, T.~D. Tillman, and G.~P. McCarthy, ``Effects of {{B}}, {{C}}, and {{Zr}} on the structure and properties of a {{P}}/{{M}} nickel base superalloy,'' {\em Metallurgical Transactions A}, vol.~18, pp.~69--77, Jan. 1987.

\bibitem{zhangSynergyPhaseMC2024}
L.~Zhang, Q.~Yang, J.~Chen, M.~Zhang, and C.~Xiao, ``Synergy of {$\gamma\prime$} phase, {{MC}} carbide and grain boundary phase on creep behavior for nickel-based superalloy {{K439B}},'' {\em Materials Science and Engineering: A}, vol.~915, p.~147261, Nov. 2024.

\bibitem{liInfluenceCarbidesPores2024a}
R.~Li, Y.~Zhang, H.~Niu, H.~Wang, and H.~Wu, ``Influence of carbides and pores on the localized deformation of nickel-based single-crystal superalloys,'' {\em Progress in Natural Science: Materials International}, vol.~34, pp.~562--568, June 2024.

\bibitem{wangDiffusionBoronAlloys1995}
W.~Wang, S.~Zhang, and X.~He, ``Diffusion of boron in alloys,'' {\em Acta Metallurgica et Materialia}, vol.~43, pp.~1693--1699, Apr. 1995.

\bibitem{keddamAssessmentBoronDiffusivities2022}
M.~Keddam and P.~Jur{\v c}i, ``Assessment of {{Boron Diffusivities}} in {{Nickel Borides}} by {{Two Mathematical Approaches}},'' {\em Materials}, vol.~15, p.~555, Jan. 2022.

\bibitem{davidFirstprinciplesStudyInsertion2020a}
M.~David, A.~Prillieux, D.~Monceau, and D.~Conn{\'e}table, ``First-principles study of the insertion and diffusion of interstitial atoms ({{H}}, {{C}}, {{N}} and {{O}}) in nickel,'' {\em Journal of Alloys and Compounds}, vol.~822, p.~153555, May 2020.

\bibitem{epifanoFirstprincipleStudySolubility2020}
E.~Epifano and G.~Hug, ``First-principle study of the solubility and diffusion of oxygen and boron in {$\gamma$}-{{TiAl}},'' {\em Computational Materials Science}, vol.~174, p.~109475, Mar. 2020.

\bibitem{zhangFirstPrinciplesCalculation2018}
X.~Zhang, X.~Li, P.~Wu, S.~Chen, S.~Zhang, N.~Chen, and X.~Huai, ``First principles calculation of boron diffusion in fcc-{{Fe}},'' {\em Current Applied Physics}, vol.~18, pp.~1108--1112, Oct. 2018.

\bibitem{jiEffectRefractoryElements2023a}
J.~Y. Ji, Z.~Zhang, J.~Chen, H.~Zhang, Y.~Z. Zhang, and H.~Lu, ``Effect of refractory elements {{M}} (={{Re}}, {{W}}, {{Mo}} or {{Ta}}) on the diffusion properties of boron in nickel-based single crystal superalloys,'' {\em Vacuum}, vol.~211, p.~111923, May 2023.

\bibitem{rajeshwarik.GrainBoundaryDiffusion2020}
S.~Rajeshwari~K., S.~Sankaran, K.~C. Hari~Kumar, H.~R{\"o}sner, M.~Peterlechner, V.~A. Esin, S.~Divinski, and G.~Wilde, ``Grain boundary diffusion and grain boundary structures of a {{Ni-Cr-Fe-}} alloy: {{Evidences}} for grain boundary phase transitions,'' {\em Acta Materialia}, vol.~195, pp.~501--518, Aug. 2020.

\bibitem{yangFirstprinciplesInvestigationInteraction2017a}
J.~Yang, J.~Huang, Z.~Ye, D.~Fan, S.~Chen, and Y.~Zhao, ``First-principles investigation on the interaction of {{Boron}} atom with nickel part {{II}}: {{Absorption}} and diffusion at grain boundary,'' {\em Journal of Alloys and Compounds}, vol.~708, pp.~1089--1095, June 2017.

\bibitem{distefanoFirstprinciplesInvestigationHydrogen2015}
D.~Di~Stefano, M.~Mrovec, and C.~Els{\"a}sser, ``First-principles investigation of hydrogen trapping and diffusion at grain boundaries in nickel,'' {\em Acta Materialia}, vol.~98, pp.~306--312, Oct. 2015.

\bibitem{heFirstprinciplesStudyHydrogen2021}
Y.~He, Y.~Su, H.~Yu, and C.~Chen, ``First-principles study of hydrogen trapping and diffusion at grain boundaries in {$\gamma$}-{{Fe}},'' {\em International Journal of Hydrogen Energy}, vol.~46, pp.~7589--7600, Feb. 2021.

\bibitem{huangHydrogenAtomSolution2023}
C.~Huang, K.~Song, S.~Zhou, Y.~Su, L.~Qiao, and L.~Gao, ``Hydrogen atom solution and diffusion behaviors at {{$\Sigma$3}} and {{$\Sigma$5}} grain boundaries of {{Fe}}, {{Ni}}, {{Cu}} and {{Al}}: {{A}} first-principles study,'' {\em Materials Today Communications}, vol.~37, p.~107222, Dec. 2023.

\bibitem{dongFastHydrogenDiffusion2017}
Y.~Dong, F.~Wang, and W.~Lai, ``Fast hydrogen diffusion along {{$\Sigma$13}} grain boundary of {$\alpha$}-{{Al2O3}} and its suppression by the dopant {{Cr}}: {{A}} first-principles study,'' {\em International Journal of Hydrogen Energy}, vol.~42, pp.~10124--10130, Apr. 2017.

\bibitem{xingNeuralNetworkKinetics2024}
B.~Xing, T.~J. Rupert, X.~Pan, and P.~Cao, ``Neural network kinetics for exploring diffusion multiplicity and chemical ordering in compositionally complex materials,'' {\em Nature Communications}, vol.~15, p.~3879, May 2024.

\bibitem{xingShortrangeOrderLocalizing2022}
B.~Xing, X.~Wang, W.~J. Bowman, and P.~Cao, ``Short-range order localizing diffusion in multi-principal element alloys,'' {\em Scripta Materialia}, vol.~210, p.~114450, Mar. 2022.

\bibitem{cadelAtomProbeTomography2002}
E.~Cadel, D.~Lemarchand, S.~Chambreland, and D.~Blavette, ``Atom probe tomography investigation of the microstructure of superalloys {{N18}},'' {\em Acta Materialia}, vol.~50, pp.~957--966, Mar. 2002.

\bibitem{tytkoMicrostructuralEvolutionNibased2012}
D.~Tytko, P.-P. Choi, J.~Kl{\"o}wer, A.~Kostka, G.~Inden, and D.~Raabe, ``Microstructural evolution of a {{Ni-based}} superalloy ({{617B}}) at 700{$^\circ$}{{C}} studied by electron microscopy and atom probe tomography,'' {\em Acta Materialia}, vol.~60, pp.~1731--1740, Feb. 2012.

\bibitem{kontisEffectBoronGrain2016a}
P.~Kontis, H.~A.~M. Yusof, S.~Pedrazzini, M.~Danaie, K.~L. Moore, P.~A.~J. Bagot, M.~P. Moody, C.~R.~M. Grovenor, and R.~C. Reed, ``On the effect of boron on grain boundary character in a new polycrystalline superalloy,'' {\em Acta Materialia}, vol.~103, pp.~688--699, Jan. 2016.

\bibitem{duPrecipitationEvolutionGrain2017a}
B.~Du, L.~Sheng, C.~Cui, J.~Yang, and X.~Sun, ``Precipitation and evolution of grain boundary boride in a nickel-based superalloy during thermal exposure,'' {\em Materials Characterization}, vol.~128, pp.~109--114, June 2017.

\bibitem{kontisRoleBoronImproving2017a}
P.~Kontis, E.~Alabort, D.~Barba, D.~M. Collins, A.~J. Wilkinson, and R.~C. Reed, ``On the role of boron on improving ductility in a new polycrystalline superalloy,'' {\em Acta Materialia}, vol.~124, pp.~489--500, Feb. 2017.

\bibitem{tekogluMetalMatrixComposite2024b}
E.~Teko{\u g}lu, A.~D. O'Brien, J.-S. Bae, K.-H. Lim, J.~Liu, S.~Kavak, Y.~Zhang, S.~Y. Kim, D.~A{\u g}ao{\u g}ullar{\i}, W.~Chen, A.~J. Hart, G.-D. Sim, and J.~Li, ``Metal matrix composite with superior ductility at 800 {$^\circ$}{{C}}: {{3D}} printed {{In718}}+{{ZrB2}} by laser powder bed fusion,'' {\em Composites Part B: Engineering}, vol.~268, p.~111052, Jan. 2024.

\bibitem{tekogluSuperiorHightemperatureMechanical2024c}
E.~Tekoglu, J.-S. Bae, H.-A. Kim, K.-H. Lim, J.~Liu, T.~Dole{\v z}al, S.~Kim, M.~Alrizqi, A.~Penn, W.~Chen, A.~Hart, J.-H. Kang, C.-S. Oh, J.~Park, F.~Sun, S.~Kim, G.-D. Sim, and J.~Li, ``Superior high-temperature mechanical properties and microstructural features of {{LPBF-printed In625-based}} metal matrix composites,'' {\em Materials Today}, vol.~80, pp.~297--307, 2024.

\bibitem{dolezalSegregationOrderingLight2025}
T.~D. Dole{\v z}al, R.~Freitas, and J.~Li, ``Segregation and ordering of light interstitials ({{B}}, {{C}}, {{H}}, and {{N}}) in {{Cr}}--{{Ni}} alloys: {{Implications}} for grain boundary stability in superalloy design,'' {\em Acta Materialia}, vol.~296, p.~121221, Sept. 2025.

\bibitem{dolezalAtomisticMechanismsOxidation2025}
T.~D. Dole{\v z}al, R.~Freitas, and J.~Li, ``Atomistic mechanisms of oxidation and chlorine corrosion in {{Ni-based}} superalloys: {{The}} role of boron and light interstitial segregation,'' {\em Acta Materialia}, vol.~301, p.~121556, Dec. 2025.

\bibitem{dolezalAtomisticSimulationsShortrange2025}
T.~D. Dole{\v z}al, E.~Tekoglu, J.-S. Bae, G.-D. Sim, R.~Freitas, and J.~Li, ``Atomistic simulations of short-range ordering with light interstitials in {{Inconel}} superalloys,'' {\em Computational Materials Science}, vol.~253, p.~113858, May 2025.

\bibitem{schuhUniversalFeaturesGrain2005a}
C.~A. Schuh, M.~Kumar, and W.~E. King, ``Universal features of grain boundary networks in {{FCC}} materials,'' {\em Journal of Materials Science}, vol.~40, pp.~847--852, Feb. 2005.

\bibitem{tschoppSymmetricAsymmetricTilt2015}
M.~A. Tschopp, S.~P. Coleman, and D.~L. McDowell, ``Symmetric and asymmetric tilt grain boundary structure and energy in {{Cu}} and {{Al}} (and transferability to other fcc metals),'' {\em Integrating Materials and Manufacturing Innovation}, vol.~4, pp.~176--189, Dec. 2015.

\bibitem{zhengGrainBoundaryProperties2020}
H.~Zheng, X.-G. Li, R.~Tran, C.~Chen, M.~Horton, D.~Winston, K.~A. Persson, and S.~P. Ong, ``Grain boundary properties of elemental metals,'' {\em Acta Materialia}, vol.~186, pp.~40--49, Mar. 2020.

\bibitem{stukowskiVisualizationAnalysisAtomistic2009a}
A.~Stukowski, ``Visualization and analysis of atomistic simulation data with {{OVITO}}--the {{Open Visualization Tool}},'' {\em Modelling and Simulation in Materials Science and Engineering}, vol.~18, p.~015012, Dec. 2009.

\bibitem{hirelAtomskToolManipulating2015}
P.~Hirel, ``Atomsk: {{A}} tool for manipulating and converting atomic data files,'' {\em Computer Physics Communications}, vol.~197, pp.~212--219, Dec. 2015.

\bibitem{rohrerComparingCalculatedMeasured2010}
G.~S. Rohrer, E.~A. Holm, A.~D. Rollett, S.~M. Foiles, J.~Li, and D.~L. Olmsted, ``Comparing calculated and measured grain boundary energies in nickel,'' {\em Acta Materialia}, vol.~58, pp.~5063--5069, Sept. 2010.

\bibitem{thompsonLAMMPSFlexibleSimulation2022a}
A.~P. Thompson, H.~M. Aktulga, R.~Berger, D.~S. Bolintineanu, W.~M. Brown, P.~S. Crozier, P.~J. Veld, A.~Kohlmeyer, S.~G. Moore, T.~D. Nguyen, R.~Shan, M.~J. Stevens, J.~Tranchida, C.~Trott, and S.~J. Plimpton, ``{{LAMMPS}} - a flexible simulation tool for particle-based materials modeling at the atomic, meso, and continuum scales,'' {\em Computer Physics Communications}, vol.~271, p.~108171, Feb. 2022.

\bibitem{takamotoUniversalNeuralNetwork2022}
S.~Takamoto, C.~Shinagawa, D.~Motoki, K.~Nakago, W.~Li, I.~Kurata, T.~Watanabe, Y.~Yayama, H.~Iriguchi, Y.~Asano, T.~Onodera, T.~Ishii, T.~Kudo, H.~Ono, R.~Sawada, R.~Ishitani, M.~Ong, T.~Yamaguchi, T.~Kataoka, A.~Hayashi, N.~Charoenphakdee, and T.~Ibuka, ``Towards universal neural network potential for material discovery applicable to arbitrary combination of 45 elements,'' {\em Nature Communications}, vol.~13, p.~2991, May 2022.

\bibitem{Matlantis}
``Matlantis, software as a service style material discovery tool.'' https://matlantis.com/.

\bibitem{henkelmanClimbingImageNudged2000}
G.~Henkelman, B.~P. Uberuaga, and H.~J{\'o}nsson, ``A climbing image nudged elastic band method for finding saddle points and minimum energy paths,'' {\em The Journal of Chemical Physics}, vol.~113, pp.~9901--9904, Dec. 2000.

\bibitem{hjorthlarsenAtomicSimulationEnvironment2017a}
A.~Hjorth~Larsen, J.~J{\o}rgen~Mortensen, J.~Blomqvist, I.~E. Castelli, R.~Christensen, M.~Du{\l}ak, J.~Friis, M.~N. Groves, B.~Hammer, C.~Hargus, E.~D. Hermes, P.~C. Jennings, P.~Bjerre~Jensen, J.~Kermode, J.~R. Kitchin, E.~Leonhard~Kolsbjerg, J.~Kubal, K.~Kaasbjerg, S.~Lysgaard, J.~Bergmann~Maronsson, T.~Maxson, T.~Olsen, L.~Pastewka, A.~Peterson, C.~Rostgaard, J.~Schi{\o}tz, O.~Sch{\"u}tt, M.~Strange, K.~S. Thygesen, T.~Vegge, L.~Vilhelmsen, M.~Walter, Z.~Zeng, and K.~W. Jacobsen, ``The atomic simulation environment---a {{Python}} library for working with atoms,'' {\em Journal of Physics: Condensed Matter}, vol.~29, p.~273002, June 2017.

\bibitem{yangFirstprinciplesInvestigationInteraction2016a}
J.~Yang, J.~Huang, D.~Fan, S.~Chen, and X.~Zhao, ``First-principles investigation on the interaction of {{Boron}} atom with {{Nickel}} part {{I}}: {{From}} surface adsorption to bulk diffusion,'' {\em Journal of Alloys and Compounds}, vol.~663, pp.~116--122, Apr. 2016.

\bibitem{chenEnhancingSulfurEmbrittlement2025a}
Y.~Chen, H.~Yu, H.~Di, and W.~Xu, ``Enhancing sulfur embrittlement resistance in {{Ni-based}} superalloys through synergistic effects of boron and carbon co-doping,'' {\em Journal of Materials Science}, vol.~60, pp.~445--462, Jan. 2025.

\bibitem{porterPhaseTransformationsMetals2021}
D.~A. Porter, K.~E. Easterling, and M.~Y. Sherif, {\em Phase {{Transformations}} in {{Metals}} and {{Alloys}}}.
\newblock Boca Raton: CRC Press, 4~ed., Nov. 2021.

\bibitem{murarkaDiffusionChromiumNickel1964}
S.~P. Murarka, M.~S. Anand, and R.~P. Agarwala, ``Diffusion of {{Chromium}} in {{Nickel}},'' {\em Journal of Applied Physics}, vol.~35, pp.~1339--1341, Apr. 1964.

\bibitem{swalinDiffusionMagnesiumSilicon1957}
R.~A. Swalin, A.~Martin, and R.~Olson, ``Diffusion of magnesium, silicon, and molybdenum in nickel,'' {\em JOM}, vol.~9, pp.~936--939, July 1957.

\bibitem{divyaInterdiffusionNiMo2010}
V.~D. Divya, S.~S.~K. Balam, U.~Ramamurty, and A.~Paul, ``Interdiffusion in the {{Ni}}--{{Mo}} system,'' {\em Scripta Materialia}, vol.~62, pp.~621--624, Apr. 2010.

\bibitem{seoEtaPhaseBoride2007a}
S.~Seo, I.~Kim, J.~Lee, C.~Jo, H.~Miyahara, and K.~Ogi, ``Eta {{Phase}} and {{Boride Formation}} in {{Directionally Solidified Ni-Base Superalloy IN792}}~+~{{Hf}},'' {\em Metallurgical and Materials Transactions A}, vol.~38, pp.~883--893, Apr. 2007.

\bibitem{guptaCompositionallyGradedNanosized2021}
R.~Gupta, K.~C.~H. Kumar, M.~J. N.~V. Prasad, and P.~Pant, ``Compositionally graded nano-sized borides in a directionally solidified nickel-base superalloy,'' {\em Scripta Materialia}, vol.~201, p.~113981, Aug. 2021.

\bibitem{gaoEffectTiB2Content2023}
J.~Gao, F.~Liu, L.~Liu, F.~Liu, Q.~You, Y.~Zheng, P.~Wang, and H.~Qiu, ``Effect of {{TiB2}} content on microstructure and mechanical properties of {{GH3536}} superalloy formed by laser solid forming,'' {\em Materials Today Communications}, vol.~37, p.~107168, Dec. 2023.

\bibitem{kitaguchiMCDecompositionBoride2024}
H.~S. Kitaguchi, L.~Small, I.~P. Jones, Y.~L. Chiu, M.~C. Hardy, and P.~Bowen, ``{{MC}} decomposition and boride formation in a next generation polycrystalline {{Ni}} based superalloy during isothermal exposure at 900~{$^\circ$}{{C}},'' {\em Materials Characterization}, vol.~209, p.~113721, Mar. 2024.

\bibitem{baeAdditiveManufacturingStrong2025a}
J.-S. Bae, E.~Tekoglu, M.~Alrizqi, A.~D. O'Brien, J.~Liu, K.~Biggs, S.~Y. Kim, A.~Penn, I.~Sulak, W.~Chen, K.~P. So, A.~J. Hart, G.-D. Sim, and J.~Li, ``Additive manufacturing of strong and ductile {{In939}}+{{TiB2}} by laser powder bed fusion,'' {\em Materials Science and Engineering: A}, vol.~939, p.~148446, Sept. 2025.

\bibitem{kassnerFundamentalsCreepMetals2009}
M.~E. Kassner, {\em Fundamentals of {{Creep}} in {{Metals}} and {{Alloys}}}.
\newblock Amsterdam London: Elsevier Science, 2009.

\bibitem{smith1948introduction}
C.~S. Smith, ``Introduction to grains, phases and interfaces -- an interpretation of microstructure,'' {\em Transactions of the Metallurgical Society of AIME}, vol.~175, pp.~15--51, 1948.

\bibitem{wagihSpectrumGrainBoundary2019}
M.~Wagih and C.~A. Schuh, ``Spectrum of grain boundary segregation energies in a polycrystal,'' {\em Acta Materialia}, vol.~181, pp.~228--237, Dec. 2019.

\bibitem{wagihLearningGrainBoundarySegregation2022}
M.~Wagih and C.~A. Schuh, ``Learning {{Grain-Boundary Segregation}}: {{From First Principles}} to {{Polycrystals}},'' {\em Physical Review Letters}, vol.~129, p.~046102, July 2022.

\bibitem{yehNanostructuredHighEntropyAlloys2004b}
J.-W. Yeh, S.-K. Chen, S.-J. Lin, J.-Y. Gan, T.-S. Chin, T.-T. Shun, C.-H. Tsau, and S.-Y. Chang, ``Nanostructured {{High-Entropy Alloys}} with {{Multiple Principal Elements}}: {{Novel Alloy Design Concepts}} and {{Outcomes}},'' {\em Advanced Engineering Materials}, vol.~6, no.~5, pp.~299--303, 2004.

\bibitem{tongMicrostructureCharacterizationAlx2005}
C.-J. Tong, Y.-L. Chen, J.-W. Yeh, S.-J. Lin, S.-K. Chen, T.-T. Shun, C.-H. Tsau, and S.-Y. Chang, ``Microstructure characterization of {{Alx CoCrCuFeNi}} high-entropy alloy system with multiprincipal elements,'' {\em Metallurgical and Materials Transactions A}, vol.~36, pp.~881--893, Apr. 2005.

\bibitem{changInfluenceSubstrateBias2008}
H.-W. Chang, P.-K. Huang, J.-W. Yeh, A.~Davison, C.-H. Tsau, and C.-C. Yang, ``Influence of substrate bias, deposition temperature and post-deposition annealing on the structure and properties of multi-principal-component ({{AlCrMoSiTi}}){{N}} coatings,'' {\em Surface and Coatings Technology}, vol.~202, pp.~3360--3366, Apr. 2008.

\bibitem{ngEntropydrivenPhaseStability2012}
C.~Ng, S.~Guo, J.~Luan, S.~Shi, and C.~T. Liu, ``Entropy-driven phase stability and slow diffusion kinetics in an {{Al0}}.{{5CoCrCuFeNi}} high entropy alloy,'' {\em Intermetallics}, vol.~31, pp.~165--172, Dec. 2012.

\bibitem{vermaDiffusionHighEntropy2024}
V.~Verma, C.~H. Belcher, D.~Apelian, and E.~J. Lavernia, ``Diffusion in {{High Entropy Alloy Systems}} -- {{{\emph{A Review}}}},'' {\em Progress in Materials Science}, vol.~142, p.~101245, Apr. 2024.

\bibitem{chenDiffusionBehaviorsFace2019}
S.~Chen, Q.~Li, J.~Zhong, F.~Xing, and L.~Zhang, ``On diffusion behaviors in face centered cubic phase of {{Al-Co-Cr-Fe-Ni-Ti}} high-entropy superalloys,'' {\em Journal of Alloys and Compounds}, vol.~791, pp.~255--264, June 2019.

\bibitem{dabrowaDemystifyingSluggishDiffusion2019}
J.~D{\k a}browa, M.~Zajusz, W.~Kucza, G.~Cie{\'s}lak, K.~Berent, T.~Czeppe, T.~Kulik, and M.~Danielewski, ``Demystifying the sluggish diffusion effect in high entropy alloys,'' {\em Journal of Alloys and Compounds}, vol.~783, pp.~193--207, Apr. 2019.

\end{thebibliography}

\end{document}